\documentclass[aps, prd, amsmath, floats, floatfix, twocolumn, nofootinbib, nosuperscriptaddress, showpacs]{revtex4-1}
\usepackage{graphicx}
\usepackage{color}
\usepackage{latexsym}
\usepackage{amsfonts}

\newcommand{\be}{\begin{eqnarray}}
\newcommand{\ee}{\end{eqnarray}}

\begin{document}

\title{Astrophysical Black Holes: A Compact Pedagogical Review}

\author{Cosimo~Bambi}
\email[E-mail: ]{bambi@fudan.edu.cn}
\affiliation{Center for Field Theory and Particle Physics and Department of Physics, Fudan University, 200433 Shanghai, China}
\affiliation{Theoretical Astrophysics, Eberhard-Karls Universit\"at T\"ubingen, 72076 T\"ubingen, Germany}

\date{\today}

\begin{abstract}
Black holes are among the most extreme objects that can be found in the Universe and an ideal laboratory for testing fundamental physics. This article will briefly review the basic properties of black holes as expected from general relativity, the main astronomical observations, and the leading astrophysical techniques to probe the strong gravity region of these objects. It is mainly intended to provide a compact introductory overview on astrophysical black holes to new students entering this research field, as well as to senior researchers working in general relativity and alternative theories of gravity and wishing to quickly learn the state of the art of astronomical observations of black holes.
\end{abstract}

\maketitle


\section{Introduction}

A {\it black hole} is, roughly speaking, a region of the spacetime in which gravity is so strong that nothing, nor even light, can escape. The {\it event horizon} is the boundary of such a region. For a more rigorous definition, see e.g.~\cite{book} and references therein, but it will not be necessary for what follows.

The possibility of the existence of extremely compact objects such that their strong gravitational field could prevent the escape of light was first discussed by John Michell and Pierre-Simon Laplace at the end of the 18th century in the context of Newtonian mechanics. In the corpuscular theory of light developed in the 17th century, light was made of small particles traveling with a finite velocity $c$. Michell and Laplace noted that the escape velocity from the surface of a body of mass $M$ and radius $R$ exceeds $c$ if $R < R_{\rm crit}$, where
\be
R_{\rm crit} = \frac{2 G_{\rm N} M}{c^2}
\ee
and $G_{\rm N}$ is Newton's gravitational constant. Such a compact body would not be able to emit radiation from its surface and should thus look black.

The theory of general relativity was proposed by Albert Einstein at the end of~1915~\cite{ein}. The simplest black hole solution was found immediately after, in 1916, by Karl Schwarzschild~\cite{sch}. It described a non-rotating black hole. However, its actual physical properties were only understood much later. David Finkelstein was the first, in~1958, to figure out that this solution had an event horizon causally separating the interior from the exterior region~\cite{fink}. The solution for a rotating black hole in general relativity was found only in 1963, by Roy Kerr~\cite{kerr}.

Even the astrophysical implications of such solutions were initially not taken very seriously. Most people were more inclined to believe that ``some unknown mechanism'' could prevent the complete collapse of a massive body and the formation of a black hole in the Universe. In 1964, Yakov ZelÕdovich and, independently, Edwin Salpeter proposed that quasars were powered by a central supermassive black hole~\cite{zeld,salp}. In the early 1970s, Thomas Bolton and, independently, Louise Webster and Paul Murdin identified the X-ray source Cygnus X-1 as the first stellar-mass black hole candidate~\cite{cyg2,cyg3}. Since then, an increasing number of astronomical observations have pointed out the existence of stellar-mass black holes in some X-ray binaries~\cite{re-mc} and of supermassive black holes at the center of many galaxies~\cite{k-r}. Thanks to technological progresses and new observational facilities, in the past 10-20~years there have been substantial progresses in the study of astrophysical black holes. In September~2015, the LIGO experiment detected, for the first time, the gravitational waves emitted from the coalescence of two black holes~\cite{gw150914}, opening a completely new window for studying these objects.

It is curious that the term black hole is relatively recent. While it is not clear who used the term first, it appeared for the first time in a publication in the January 18, 1964 issue of Science News Letter. It was on a report on a meeting of the American Association for the Advancement of Science by journalist Ann Ewing. The term became quickly very popular after it was used by John Wheeler at a lecture in New York in 1967.

Black holes can potentially have any value of the mass, and the latter is the characteristic quantity setting the size of the system. The {\it gravitational radius} of an object of mass $M$ is defined as
\be
r_{\rm g} &=& \frac{G_{\rm N} M}{c^2} 
= 14.77 \left( \frac{M}{10 \; M_\odot} \right) \text{ km } \, .
\ee
The associated characteristic time scale is
\be
\tau &=& \frac{r_{\rm g}}{c}
= 49.23 \left( \frac{M}{10 \; M_\odot} \right) \text{ $\mu$s } \, .
\ee
It can be quite useful to have these two scales in mind. For $M \sim 10^6$~$M_\odot$, we find $r_{\rm g} \sim 10^6$~km and $\tau \sim 5$~s. For $M \sim 10^9$~$M_\odot$, we have $r_{\rm g} \sim 10^9$~km and $\tau \sim 1$~hr.

When we discuss observations of astrophysical black holes, an important concept is that of {\it Eddington luminosity}. It is the maximum luminosity for a generic object, not necessarily a black hole. The Eddington luminosity $L_{\rm Edd}$ is reached when the pressure of the radiation luminosity on the emitting material balances the gravitational force towards the object. If a normal star has a luminosity $L > L_{\rm Edd}$, the pressure of the radiation luminosity drives an outflow. If the luminosity of the accretion flow of a black hole exceeds $L_{\rm Edd}$, the pressure of the radiation luminosity stops the accretion process, reducing the luminosity. Assuming that the emitting medium is a ionized gas of protons and electrons, the Eddington luminosity of an object of mass $M$ is
\be
L_{\rm Edd} &=& \frac{4 \pi G_{\rm N} M m_p c}{\sigma_{\rm Th}} \nonumber\\
&=& 1.26 \cdot 10^{38} \left(\frac{M}{M_\odot}\right) \text{ erg/s} \, ,
\ee
where $m_p$ is the proton mass and $\sigma_{\rm Th}$ is the electron Thomson cross section.  For an accreting black hole, we can define the Eddington mass accretion rate $\dot{M}_{\rm Edd}$ from $L_{\rm Edd} = \eta_{\rm r} \dot{M}_{\rm Edd} c^2$, where $\eta_{\rm r} \sim 0.1$ is the radiative efficiency of the accretion process, namely the fraction of energy of the accreting material emitted in the form of electromagnetic radiation.


\section{Black Holes in General Relativity}

In 4-dimensional general relativity, black holes are relatively simple objects, in the sense they are completely characterized by a small number of parameters. This is the result of the {\it no-hair theorem}, which holds under specific assumptions~\cite{h1,h2,h3,h4}. The name no-hair is to indicate that black holes have only a small number of features (hairs). We have also a {\it uniqueness theorem}, according to which black holes are only characterized by a family of solutions. Violations of these theorems are possible if we relax some of these assumptions or we consider theories beyond general relativity.

A {\it Schwarzschild black hole} is a non-rotating and electrically uncharged black hole and is completely characterized by one parameter, the black hole mass $M$. A {\it Reissner-Nordstr\"om black hole} is a non-rotating black hole of mass $M$ and electric charge $Q$. A {\it Kerr black hole} is an uncharged black hole of mass $M$ and spin angular momentum $J$. The general case is represented by a {\it Kerr-Newman black hole}, which has a mass $M$, a spin angular momentum $J$, and an electric charge $Q$.

In what follows, we will only consider Kerr black holes (which include the Schwarzschild case for vanishing spin angular momentum), because for astrophysical macroscopic objects the possible non-vanishing electric charge is extremely small and can be ignored~\cite{book}. Instead of $M$ and $J$, it is often convenient to use $M$ and $a_*$, where $a_*$ is the dimensionless spin parameter
\be
a_* = \frac{c J}{G_{\rm N} M^2} \, .
\ee 
Note that in Newtonian gravity the spin does not play any role; in Newton's Universal Law of Gravitation we have only the masses of the bodies, not their spins. This is not true in general relativity.

In general relativity, the choice of the coordinate system is arbitrary, and therefore the numerical values of the coordinates have no physical meaning. Despite that, it can be useful to know some quantities in certain coordinates. The Boyer-Lindquist coordinates are quite commonly used to describe Kerr black holes. In this coordinate system, the radial coordinate of the event horizon of a Kerr black hole is
\be\label{eq-horizon}
r_{\rm H} = r_{\rm g} \left( 1 + \sqrt{1 - a_*^2} \right) \, ,
\ee
and 
ranges from $2 \, r_{\rm g}$ for a non-rotating black hole ($a_* = 0$, Schwarzschild black hole) to $r_{\rm g}$ for a maximally rotating black hole ($a_* = \pm 1$). As we can see from Eq.~(\ref{eq-horizon}), the spin parameter is subject to the constraint $| a_* | \le 1$ ({\it Kerr bound}). For $| a_* | > 1$, the Kerr solution has no horizon, and instead of a black hole we have a naked singularity. In what follows, we will ignore such a possibility.

If we consider the motion of a test-particle around a point-like massive body in Newtonian gravity, equatorial circular orbits (i.e. orbits in the plane perpendicular to the spin of the object) are always stable. However, this is not true for a test-particle orbiting a Kerr black hole, and we have the existence of an {\it innermost stable circular orbit}, more often called ISCO. In Boyer-Lindquist coordinates, the ISCO radius is $6 \, r_{\rm g}$ for a Schwarzschild black hole and move to $r_{\rm g}$ ($9 \, r_{\rm g}$) for a maximally rotating black hole and a corotating (counterrotating) orbit, namely an orbit with angular momentum parallel (antiparallel) to the black hole spin.

\begin{figure}[t]
\begin{center}
\includegraphics[type=pdf,ext=.pdf,read=.pdf,width=8.7cm]{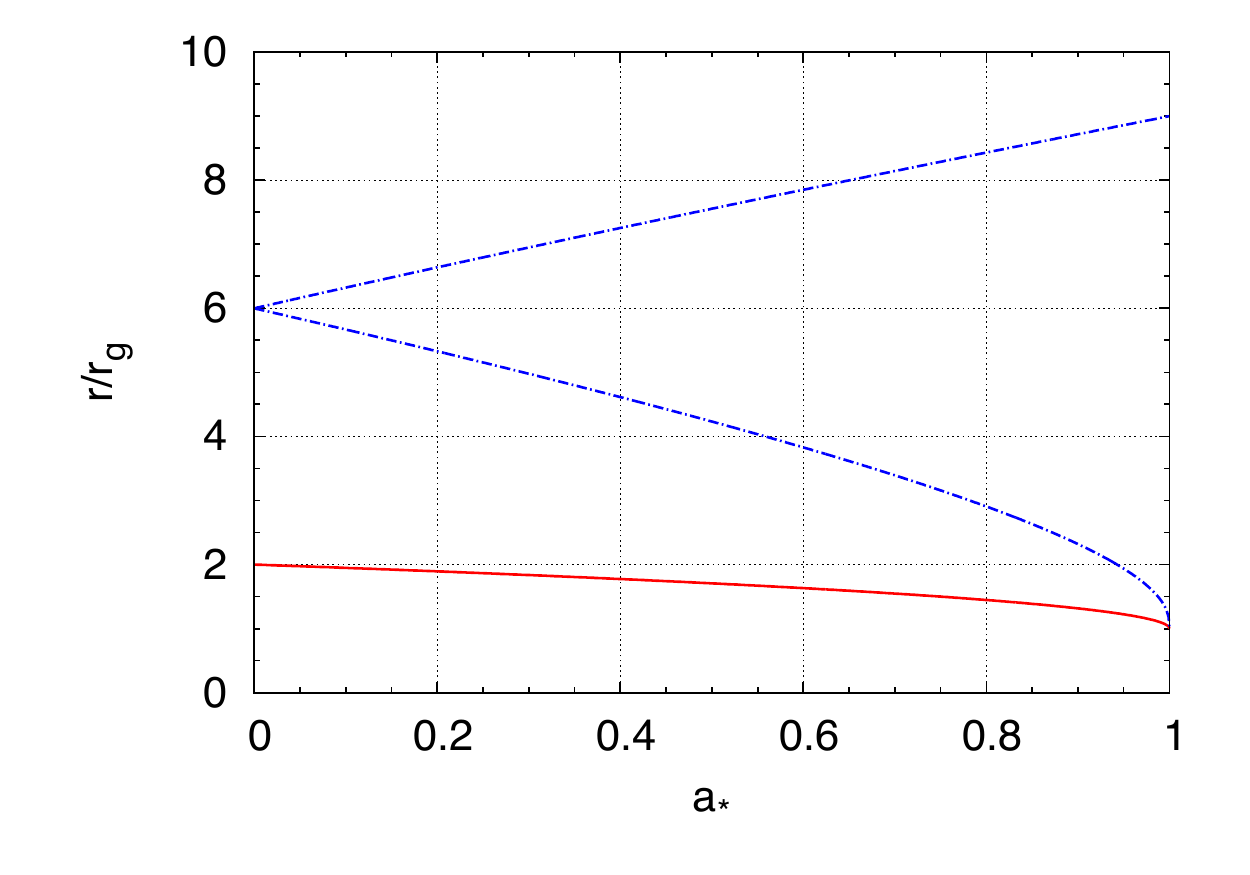}
\end{center}
\vspace{-0.6cm}
\caption{Radius of the event horizon (red solid line) and of the ISCO (blue dash-dotted line) of a Kerr black hole in Boyer-Lindquist coordinates as a function of the spin parameter $a_*$. For the ISCO radius, the upper curve refers to counterrotating orbits and the lower curve to corotating orbits. \label{f-isco}}
\end{figure}

\begin{table}[h]
\centering
\begin{tabular}{ccccccc}
\hline
$a_*$ & \hspace{0.5cm} & $r_{\rm H}/r_{\rm g}$ & \hspace{0.5cm} & $r_{\rm ISCO}/r_{\rm g}$ & \hspace{0.5cm} & $\eta_{\rm NT}$ \\
\hline
-1 && 1 && 9 && 0.038 \\
0 && 2 && 6 && 0.057 \\
0.5 && 1.866 && 4.423 && 0.082 \\
0.8 && 1.6 && 3.065 && 0.121 \\
0.9 && 1.436 && 2.424 && 0.155 \\
0.95 && 1.312 && 2.000 && 0.190 \\
0.99 && 1.141 && 1.474 && 0.264 \\
0.998 && 1.063 && 1.243 && 0.321 \\
1 && 1 && 1 && 0.423 \\
\hline
\end{tabular}
\vspace{0.1cm}
\caption{Properties of Kerr black holes. For every spin parameters $a_*$, the table shows the corresponding radius of the event horizon $r_{\rm H}$, the radius of the ISCO $r_{\rm ISCO}$, and the radiative efficiency of a Novikov-Thorne disk $\eta_{\rm NT}$ [see Section~\ref{ss-nt}, where $\eta_{\rm NT}$ is defined in Eq.~(\ref{eq-thin-ntradeff})]. $r_{\rm H}$ and $r_{\rm ISCO}$ in Boyer-Lindquist coordinates. $a_*>0$ $(<0)$ for corotating (counterrotating) orbits. \label{t-k}}
\end{table}

Fig.~\ref{f-isco} shows the radial values of the event horizon $r_{\rm H}$ and of the ISCO radius $r_{\rm ISCO}$ in Boyer-Lindquist coordinates as a function of the black hole spin parameter $a_*$. Table~\ref{t-k} reports some numerical values of $r_{\rm H}$, $r_{\rm ISCO}$, and $\eta_{\rm NT}$ (see Section~\ref{ss-nt}) for specific values of $a_*$.

More details on black holes in general relativity can be found in~\cite{book,chandra,mtw}.


\section{Astrophysical Black Holes}

From general relativity, there are no constraints on the value of the mass of a black hole, which can thus be arbitrarily small as well as arbitrarily large. From astronomical observations, we have strong evidence of at least two classes of astrophysical black holes:
\begin{enumerate}
\item {\it Stellar-mass black holes}~\cite{re-mc}.
\item {\it Supermassive black holes}~\cite{k-r}.
\end{enumerate}
There is also some evidence of {\it intermediate-mass black holes}, with a mass filling the gap between the stellar-mass and the supermassive ones~\cite{inter}. Black holes should form from the complete gravitational collapse of a system, when there is no mechanics capable of balancing the gravitational force and the system shrinks until the formation of the event horizon. The collapse of the core of heavy stars is expected to produce black holes with a mass $M \gtrsim 3$~$M_\odot$ because for cores of lower mass the quantum pressure of neutrons should stop the collapse and the final product should be a neutron star~\cite{bh1,bh2,bh3}. However, there are cosmological scenarios in which it is possible to produce {\it primordial black holes} with any mass, even much lower than 3~$M_\odot$~\cite{primordial}. Nevertheless, for the moment there is no evidence for the existence of such objects.

Note the different terminology employed in different scientific communities. Among astronomers, it is common to call ``black hole'' an astrophysical object that is supposed to be a black hole and for which there is a dynamical measurement of its mass. The latter indeed guarantees that the object is (if it is compact) too heavy for being a neutron star. ``Black hole candidates'' are instead astrophysical objects that are supposed to be black holes but for which there is no dynamical measurement of their mass. In the theoretical physics community, every astrophysical object that is supposed to be a black hole is called ``black hole candidate'' because it is only possible to put some constraints on the existence of the event horizon, but it is impossible to get a proof.

\subsection{Stellar-Mass Black Holes}

From stellar evolution simulations, we expect that in our Galaxy there is a population of about $10^8-10^9$~black holes formed at the end of the evolution of heavy stars~\cite{bhnum0,bhnum}, and the same number can be expected in similar galaxies. The initial mass of a stellar-mass black hole should depend on the properties of the progenitor star: on its mass, its evolution, and the supernova explosion mechanism~\cite{mass}. A crucial quantity is the metallicity of the star, namely the fraction of mass of the star made of elements heavier than helium.

The maximum mass of black hole remnants critically depends on the metallicity. The final mass of the remnant is indeed determined by the mass loss rate by stellar winds, which increases with the metallicity because heavier elements have a larger cross section than lighter ones, and therefore they evaporate faster. For a low-metallicity star~\cite{gg1,gg2,gg3}, there may be a mass gap in the remnant, roughly between 50 and 150~$M_\odot$, namely the mass of the black hole remnant can be $M \lesssim 50$~$M_\odot$ or $M \gtrsim 150$~$M_\odot$. As the metallicity increases, black holes with $M \gtrsim 150$~$M_\odot$ disappear, mainly because of the increased mass loss rate. Note, however, that some models do not find remnants with a mass above the gap, because stars with $M \gtrsim 150$~$M_\odot$ may undergo a runaway thermonuclear explosion that completely destroys the system, without leaving any black hole remnant~\cite{gg1,gg2}.

The lower bound may come from the maximum mass for a neutron star: the exact value is currently unknown, because it depends on the equation of state of matter at super-nuclear densities, but it should be around $2-3$~$M_\odot$. For bodies with a mass lower than this limit, the quantum neutron pressure can stop the collapse and the final product is a neutron stars. For bodies exceeding this limit, the final product is a black hole~\cite{bh1,bh2,bh3}. Note, however, that there may be a mass gap between the maximum neutron star mass and the minimum black hole mass~\cite{mass-gap}.

Stellar-mass black holes may thus have a mass in the range $3-100$~$M_\odot$. At the moment, all the known stellar-mass black holes in X-ray binaries have a mass $M \approx 3 - 20$~$M_\odot$~\cite{mass2}. Gravitational waves have shown the existence of heavier stellar-mass black holes. In particular, the event called GW150914 was associated to the coalescence of two black holes with a mass $M \approx 30$~$M_\odot$ that merged to form a black hole with $M \approx 60$~$M_\odot$~\cite{gw150914}.

While we expect a huge number of stellar-mass black holes in the Galaxy, we only know about 20~objects with a dynamical measurement of the mass and about 50~objects without a dynamical measurement of their mass (it is thus possible that some of them are not black holes but neutron stars). This is because their detection is very challenging. The simplest case is when the black hole is in a binary system and has a companion star. The presence of a compact object can be discovered from the observation of X-ray radiation emitted from the inner part of the accretion disk (see Section~\ref{s-ad} for more details about accretion). If we can study the orbital motion of the companion star, we may be able to measure the mass function~\cite{mass2}
\be
f (M) = \frac{K_{\rm c}^3 P_{\rm orb}}{2 \pi G_{\rm N}} 
= \frac{M \sin^3 i}{\left( 1 + q \right)^2} \, ,
\ee
where $K_{\rm c} = v_{\rm c} \sin i$, $v_{\rm c}$ is the velocity of the companion star, $i$ is the angle between the normal of the orbital plane and our line of sight, $P_{\rm orb}$ is the orbital period of the system, $q = M_{\rm c}/M$, $M_{\rm c}$ is the mass of the companion, and $M$ is the mass of the dark object. If we can somehow estimate $i$ and $M_{\rm c}$, we can infer $M$, and in this case we talk about dynamical measurement of the mass. The dark object is a black hole if $M > 3$~$M_\odot$~\cite{bh1,bh2,bh3}.

\begin{figure}[t]
\begin{center}
\includegraphics[type=pdf,ext=.pdf,read=.pdf,width=8.7cm]{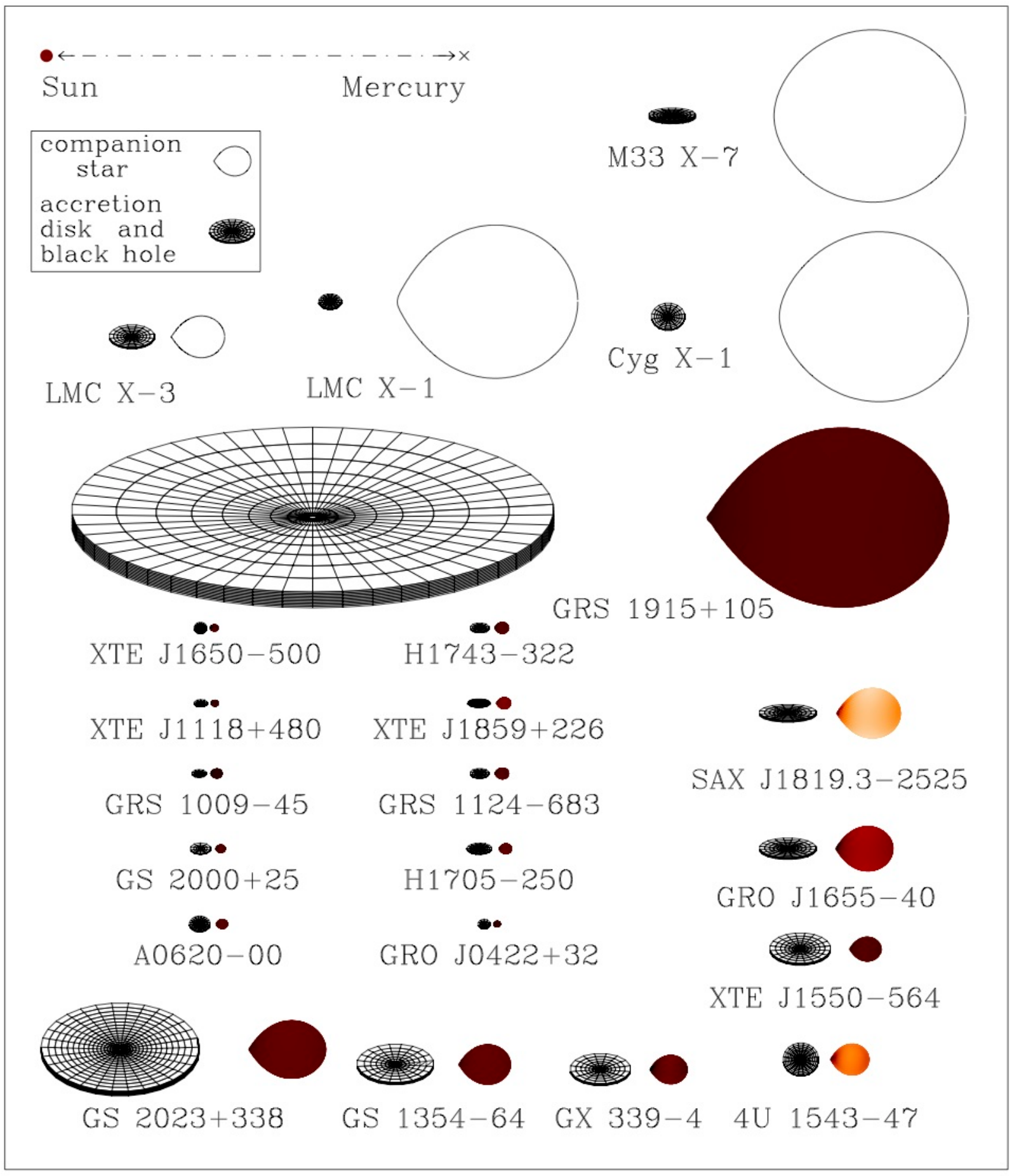}
\end{center}
\vspace{-0.3cm}
\caption{Sketch of 22~X-ray binaries with a stellar-mass black hole confirmed by dynamical measurements. For every system, the black hole accretion disk is on the left and the companion star is on the right. The color of the companion star roughly indicates its surface temperature (from brown to white as the temperature increases). The orientation of the disks indicates the inclination angles of the binaries. For comparison, in the top left corner of the figure we see the system Sun-Mercury: the distance between the two bodies is about 50~millions km and the radius of the Sun is about 0.7~millions km. Figure courtesy of Jerome Orosz. \label{f-bhb}}
\end{figure}

Black holes in X-ray binaries (black hole binaries\footnote{Generally speaking, a {\it black hole binary} is a binary system in which at least one of the two bodies is a black hole, and a {\it binary black hole} is a binary system of two black holes. In the context of stellar-mass black holes, the term black hole binary is used to indicate a binary system of a black hole with a stellar companion. In the context of supermassive black holes, it is common to call black hole binary a system of two supermassive black holes.}) are grouped into two classes: {\it low-mass X-ray binaries} (LMXBs) and {\it high-mass X-ray binaries} (HMXBs). Low and high is referred to the stellar companion, not to the black hole: in the case of LMXBs, the companion star has normally a mass $M < 3$~$M_\odot$, while for HMXBs the companion star has $M > 10$~$M_\odot$. Observationally, we can classify black hole binaries either as {\it transient X-ray sources} or {\it persistent X-ray sources}. LMXBs are usually transient sources, because the mass transfer is not continuos (for instance, at some point the surface of the companion star may expand and the black hole strips some gas): the system may be bright for a period ranging from some days to a few months and then be in a quiescent state for months or even decades. We expect $10^3 - 10^4$~LMXBs in the Galaxy~\cite{lm1,lm2} and every year we discover 1-2~new objects, when they pass from their quiescent state to an outburst (see Section~\ref{ss-ss} for more details). HMXBs are persistent sources: the mass transfer from the companion star to the black hole is a relatively regular process (typically it is due to the stellar wind of the companion) and the binary is a bright source at any time without quiescent periods.

Fig.~\ref{f-bhb} shows 22~X-ray binaries with a stellar-mass black hole confirmed by dynamical measurements. To have an idea of the size of these systems, the figure also shows the Sun (whose radius is 0.7~millions km) and the distance Sun-Mercury (about 50~millions km). The black holes have a radius $< 100$~km and cannot be seen, but we can clearly see their accretion disks formed from the transfer of material from the companion star. The latter may have a quite deformed shape (in particular, we can see some cusps) due to the the tidal force produced by the gravitational field of the black hole. In Fig.~\ref{f-bhb}, Cygnus~X-1 (Cyg~X-1 in Fig.~\ref{f-bhb}), LMC~X-1, LMC~X-3, and M33~X-7 are HMXBs, while all other systems are LMXBs. Among these HMXBs, only Cygnus~X-1 is in our Galaxy. Among the LMXBs, there is GRS~1915+105, which is quite a peculiar source: since 1992, it is a bright X-ray source in the sky, so it can be considered a persistent source even if it is a LMXB. This is probably because of its large accretion disk, which can provide enough material at any time.

\begin{figure}[b]
\begin{center}
\includegraphics[type=pdf,ext=.pdf,read=.pdf,width=8.7cm]{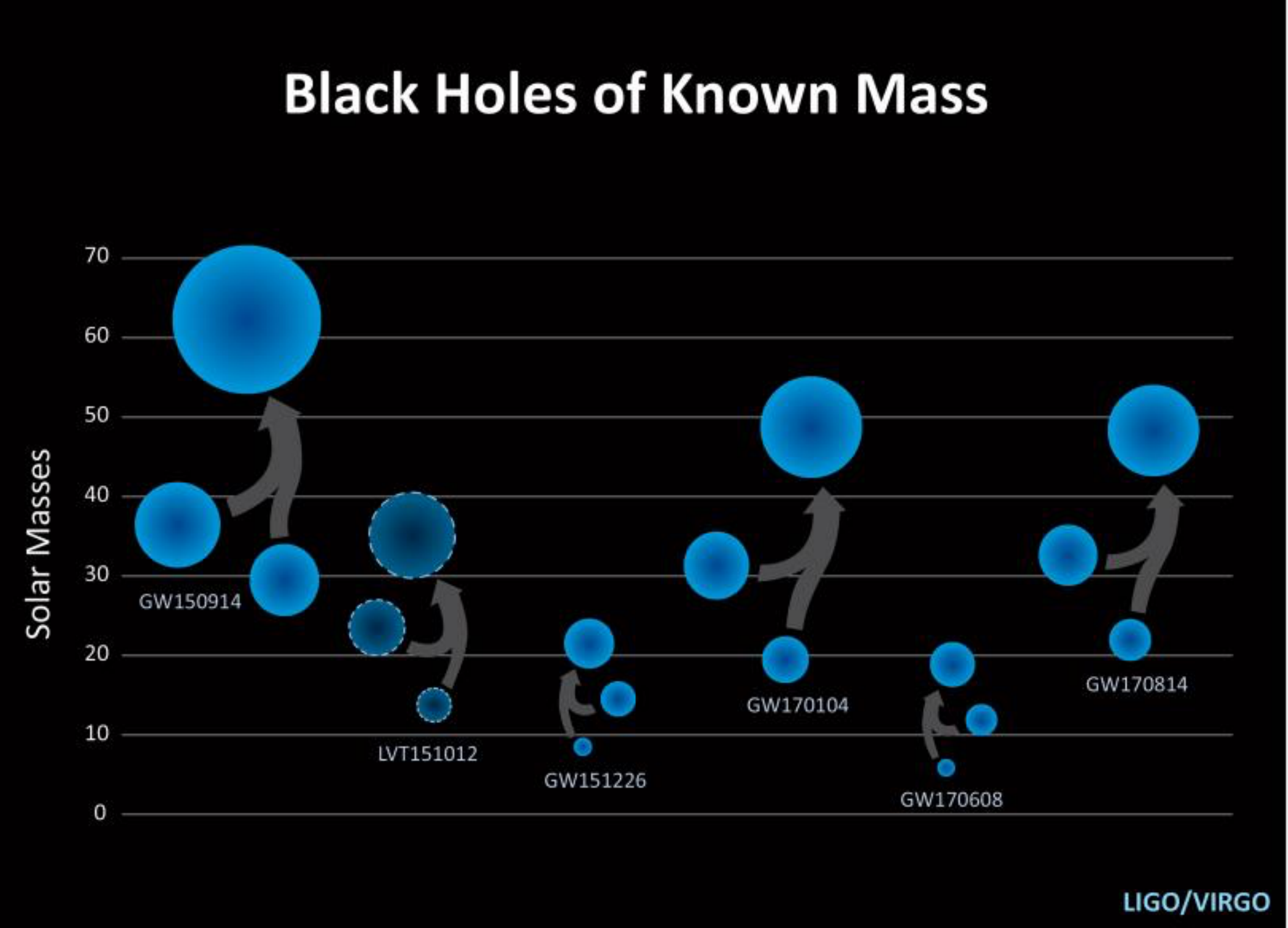}
\end{center}
\vspace{-0.3cm}
\caption{Masses of the first black holes observed with gravitational waves, with the two initial objects merging into a larger one, as shown by the arrows. Image Credit: LIGO/NSF/Caltech/SSU Aurore Simmonet.  \label{f-bhb2}}
\end{figure}

Black holes in compact binary systems (black hole-black hole or black hole-neutron star) can be detected with gravitational waves when the signal is sufficiently strong, which means just before the merger (see Section~\ref{ss-gw} for more details). Fig.~\ref{f-bhb2} shows the first detections by the LIGO/Virgo collaboration. The name of the event is classified as GW (gravitational wave event) and then there is the date: for example, GW150914 was detected on 14~September~2015. LVT151012 is not classified as a gravitational wave event because it may have been caused by noise. For every event, the figure shows the two original black holes as well as the final one after merger.

Isolated black holes are much more elusive. In principle, they can be detected by observing the modulation of the light of background stars due to the gravitational lensing caused by the passage of a black hole along the line of sign of the observer~\cite{lensing}.

\subsection{Supermassive Black Holes}

Astronomical observations show that at the center of a large number of galaxies there is a large amount of mass in a relatively small volume. The standard interpretation is that these objects are supermassive black holes with $M \sim 10^5 - 10^{10}$~$M_\odot$. The strongest constraints come from the center of our Galaxy\footnote{The total mass of our Galaxy is estimated to be $M_{\rm MW} \sim 10^{12}$~$M_\odot$~\cite{mass-g}. The mass of the central supermassive black hole is $M \approx 4 \cdot 10^6$~$M_\odot$~\cite{mass-bmw}.} and of NGC~4258 by studying the motion of individual stars or of gas in their nuclei. In the end, we can exclude the existence of a cluster of compact non-luminous bodies like neutron stars and therefore we can conclude that these objects are supermassive black holes~\cite{maoz}. In the case of other galaxies, it is not possible to put such constraints with the available data, but it is thought that every middle-size (like our Galaxy) or large galaxy has a supermassive black hole at its center\footnote{Exceptions may be possible: the galaxy A2261-BCG has a very large mass but it might not have any supermassive black hole at its center~\cite{postman}.}. For lighter galaxies, the situation is more uncertain. Most models predict supermassive black holes at the center of lighter galaxies as well~\cite{mv}, but there are also models predicting the existence of a population of faint low-mass galaxies with no supermassive black hole at their center~\cite{mv07a,mv07b}. Observations suggest that some small galaxies have a supermassive black hole and other small galaxies do not~\cite{ferrarese,amuse}.

In the case of stellar-mass black holes, it is easy to argue that they are the final product of the evolution of very heavy stars. In the case of supermassive black holes, at the moment we do not know their exact origin. We observe supermassive objects in galactic nuclei with a mass $M \sim 10^5 - 10^{10}$~$M_\odot$. More puzzlingly, we observe objects with masses $M \sim 10^{10}$~$M_\odot$ even in very distant galaxies~\cite{wu15}, when the Universe was only 1~billion years old, and we do not know how such objects were created and were able to grow so fast in a relatively short time~\cite{mv}. The Eddington accretion rate can be exceeded in some accretion models, and this may indeed be a possible path to the rapid growth of supermassive black holes~\cite{super-e1}. The possibility of super-Eddington accretion is confirmed, for instance, by the observation of a neutron star in the galaxy M82 with a luminosity exceeding its Eddington limit~\cite{super-e2}. It is also possible that supermassive black holes formed from the collapse of heavy primordial clouds rather than of stars, or that they formed from the merger of several black holes~\cite{mv}.

\subsection{Intermediate-Mass Black Holes}

Intermediate-mass black holes are, by definition, black holes with a mass between the stellar-mass and the supermassive ones, say $M \sim 10^2 - 10^4$~$M_\odot$. At the moment, there is no dynamical measurement of the mass of these objects, and their actual nature is still controversial.

Some intermediate-mass black hole candidates are associated to ultra luminous X-ray sources~\cite{ulxs}. These objects have an X-ray luminosity $L_X > 10^{39}$~erg/s, which exceeds the Eddington luminosity of a stellar-mass object, and they may thus have a mass in the range $10^2 - 10^4$~$M_\odot$. However, we cannot exclude they are actually stellar-mass black holes (or neutron stars~\cite{super-e2}) with non-isotropic emission and a moderate super-Eddington mass accretion rate~\cite{mine}.

The existence of intermediate-mass black holes is also suggested by the detection of some quasi-periodic oscillations (QPOs, see Section~\ref{ss-qpos}) in some ultra-luminous X-ray sources. QPOs are currently not well understood, but they are thought to be associated to the fundamental frequencies of the oscillation of a particle around a black hole. Since the size of the system scales as the black hole mass, QPOs should scale as $1/M$, and some observations may indicate the existence of compact objects with masses in the range $10^2 - 10^4$~$M_\odot$~\cite{i-qpo}.

Intermediate-mass black holes may be expected to form at the center of dense stellar clusters by merger. Several studies have tried to explore the possible existence of these objects from the observations of the motion of the stars in certain clusters. The presence of an intermediate-mass black hole at the center of the cluster should increase the velocity dispersion in the cluster. Some studies suggest that there are indeed intermediate-mass black holes at the center of certain globular cluster~\cite{clu1,clu2}, but there is not yet a common consensus.


\section{Accretion Disks \label{s-ad}}

A black hole itself cannot emit any radiation by definition. On the contrary, we can observe the radiation emitted by the gas in a possible accretion disk surrounding the black hole. In the case of stellar-mass black holes with a companion star, the disk is created by the mass transfer from the stellar companion to the black hole. In the case of supermassive black holes in galactic nuclei, the disk forms from the material in the interstellar medium~\cite{fuel1} or as a result of galaxy merger~\cite{fuel2,fuel3}.

The accretion disk can have different shapes and different properties, depending on its exact origin. An accretion disk is {\it geometrically thin} ({\it thick}) if $h/r \ll 1$ ($h/r \sim 1$), where $h$ is the semi-thickness of the disk at the radial coordinate $r$. The disk is {\it optically thin} ({\it thick}) if $h \ll \lambda$ ($h \gg \lambda$), where $\lambda$ is the photon mean free path in the medium of the disk. If the disk is optically thick, we see the radiation emitted from the surface of the disk, like in the case of stars.

An important class of accretion disks is represented by the geometrically thin and optically thick disks, which are commonly described by the Novikov-Thorne model~\cite{ntm,ntm2}.

\subsection{Novikov-Thorne Disks \label{ss-nt}}

The Novikov-Thorne model is the standard framework for the description of geometrically thin and optically thick accretion disks around black holes. The main assumptions of the model are:
\begin{enumerate}
\item The accretion disk is geometrically thin ($h/r \ll 1$).
\item The accretion disk is perpendicular to the black hole spin.
\item The inner edge of the disk is at the ISCO radius.
\item The motion of the particle gas in the disk is determined by the gravitational field of the black hole, while the impact of the gas pressure is ignored. 
\end{enumerate}
For the full list of assumptions and a detailed discussion, see e.g.~\cite{book,ntm2} and references therein. Here we just note that the assumption~2 can be realized by the Bardeen-Petterson effect~\cite{bp1,bp2,bp3}, which is the combination of the relativistic precession of the disk with its viscosity and drags the innermost part of the disk to align the disk angular momentum with the black hole spin.

The accretion process in the Novikov-Thorne model can be summarized as follows. The particles of the accreting gas slowly fall onto the central black hole. When they reach the ISCO radius, they quickly plunge onto the black hole without emitting additional radiation. The total power of the accretion process is $L_{\rm acc} = \eta \dot{M} c^2$, where $\eta = \eta_{\rm r} + \eta_{\rm k}$ is the total efficiency, $\eta_{\rm r}$ is the radiative efficiency, and $\eta_{\rm k}$ is the fraction of gravitational energy converted to kinetic energy of jets/outflows. The Novikov-Thorne model assumes that $\eta_{\rm k}$ can be ignored, and therefore the radiative efficiency of a Novokov-Thorne accretion disk is
\be\label{eq-thin-ntradeff}
\eta_{\rm NT} = 1 - E_{\rm ISCO} \, ,
\ee 
where $E_{\rm ISCO}$ is the energy per unit rest-mass of the gas at the ISCO radius~\cite{book}. The fourth column in Table~\ref{t-k} shows the Novikov-Thorne radiative efficiency for specific values of $a_*$. Note that the accretion process onto a black hole is an extremely efficient mechanism to convert mass into energy. If we consider the nuclear reactions inside stars, their efficiency is less than 1\%. In the case of the Novikov-Thorne accretion process, the efficiency is 5.7\% for a Schwarzschild black hole, and increases for higher spins and corotating disks up to 42\% for a maximally rotating Kerr black hole.

\subsection{Evolution of the spin parameter \label{ss-evolution}}

An accreting black hole changes its mass $M$ and spin angular momentum $J$ as it swallows more and more material from its disk. In the case of a Novikov-Thorne disk, it is relatively easy to calculate the evolution of these parameters. If we assume that the gas in the disk emits radiation until it reaches the ISCO radius and then quickly plunges onto the black hole, the evolution of the spin parameter $a_*$ is governed by the following equation~\cite{thorne}
\be\label{eq-evolution}
\frac{d a_*}{d \ln M} = \frac{c}{r_{\rm g}} 
\frac{L_{\rm ISCO}}{E_{\rm ISCO}} - 2 a_* \, ,
\ee
where $L_{\rm ISCO}$ is the angular momentum per unit rest-mass of the gas at the ISCO radius. In the Kerr metric, assuming an initially non-rotating black hole of mass $M_0$, the solution of Eq.~(\ref{eq-evolution}) is
\be
\hspace{-0.6cm}
a_* =  
\begin{cases}
\sqrt{\frac{2}{3}}
\frac{M_0}{M} \left[4 - \sqrt{18 \, \frac{M_0^2}{M^2} - 2}\right] 
 & \text{if } M \le \sqrt{6} \, M_0 \, , \\
1 & \text{if } M > \sqrt{6} \, M_0 \, .
\end{cases}
\ee
The black hole spin parameter $a_*$ monotonically increases from 0 to 1 and then remain constant. $a_* = 1$ is the equilibrium spin parameter and is reached after the black hole has increased its mass by the factor $\sqrt{6} \approx 2.4$.

If we take into account the fact that the gas in the accretion disk emits radiation and that a fraction of this radiation is captured by the black hole, Eq.~(\ref{eq-evolution}) becomes
\be
\frac{d a_*}{d \ln M} = \frac{c}{r_{\rm g}} 
\frac{L_{\rm ISCO} + \zeta_{\rm L}}{E_{\rm ISCO} + \zeta_{\rm E}} - 2 a_* \, ,
\ee
where $\zeta_{\rm L}$ and $\zeta_{\rm E}$ are related to the amount of photons captured by the black holes and must be computed numerically. Now the equilibrium value of the spin parameter is not 1 but the so-called {\it Thorne limit} $a_*^{\rm Th} \approx 0.998$ (its exact numerical value depends on the emission properties of the gas in the disk)~\cite{thorne}.

In the case of stellar-mass black holes in X-ray binaries, the spin should not change much from its original value~\cite{sp1}; see, however, \cite{sp2}. If the black hole is in a LMXB, the mass of the companion is a small fraction with respect to that of the black hole, and therefore the black hole cannot substantially change its mass and spin even after swallowing the whole companion star. If the black hole is in a HMXB, the stellar companion has a lifetime too short to transfer enough material to the black hole even assuming a mass accretion rate at the Eddington limit.

The situation is different in the case of supermassive black holes. In the case of prolonged disk accretion, the object may indeed get a very high spin, possibly close to the Thorne limit. However, there may be other events to challenge it. For example, in the case of galaxy mergers the black holes in their nuclei should merge too, and the final product is unlikely a black hole of high spin~\cite{sp3}. Even accretion from randomly distributed bodies may spin the black hole down~\cite{sp4,sp5}. More details on the spin evolution of supermassive black holes can be found, for instance, in~\cite{sss1,sss2,sss3,sss4}.


\section{Electromagnetic Spectrum}

\begin{figure}[t]
\begin{center}
\includegraphics[type=pdf,ext=.pdf,read=.pdf,width=8.7cm]{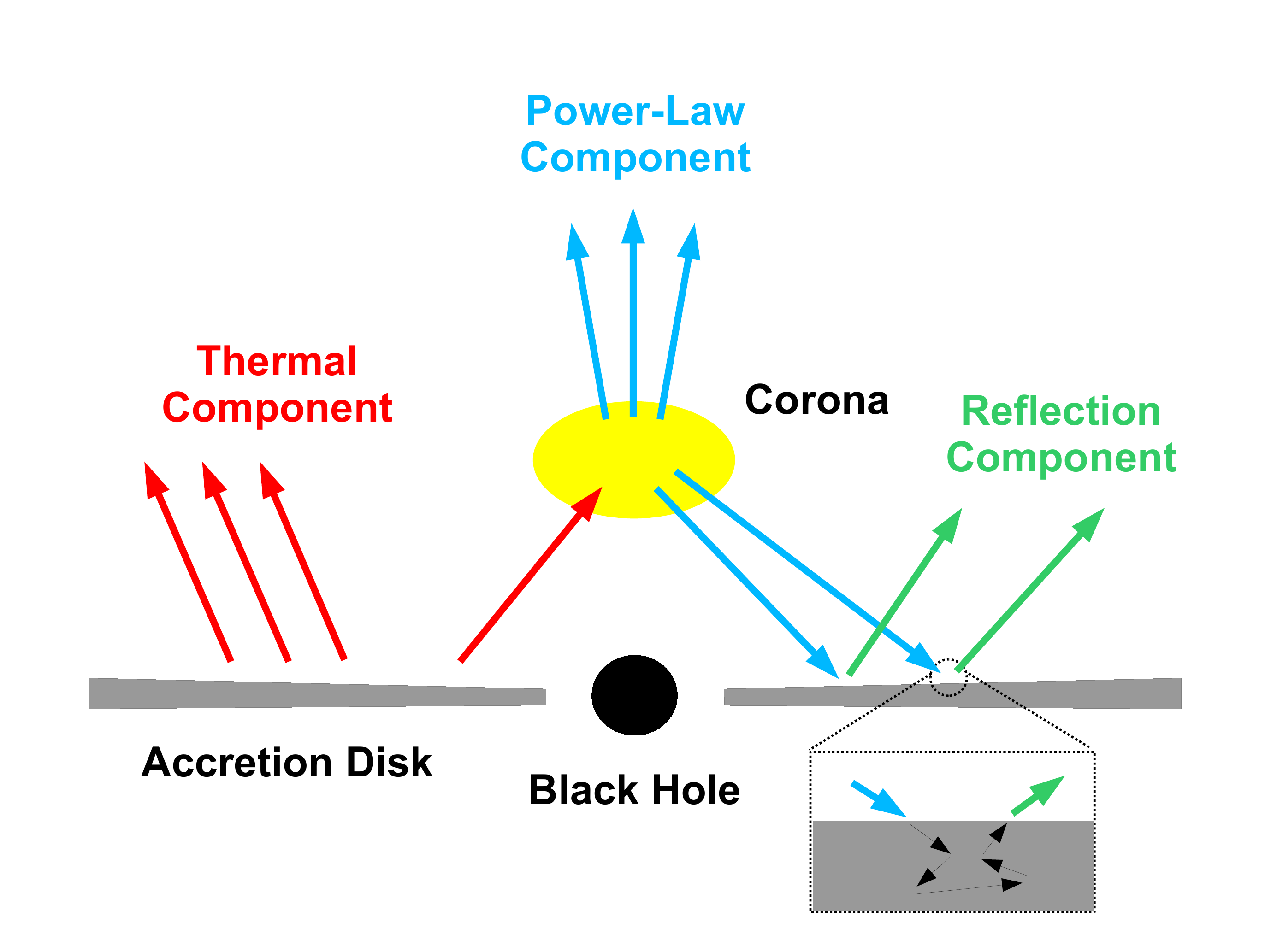}
\end{center}
\vspace{-0.3cm}
\caption{Disk-corona model. The black hole is surrounded by a thin accretion disk with a multi-color blackbody spectrum (red arrows). Some thermal photons from the disk have inverse Compton scattering off free electrons in the corona, producing a power-law component (blue arrows). The latter also illuminates the disk, generating a reflection component (green arrows). \label{f-corona}}
\end{figure}

In the disk-corona model, a black hole accretes from a geometrically thin and optically thick accretion disk, see Fig.~\ref{f-corona}. The disk emits as a blackbody locally and as a multi-color blackbody when integrated radially\footnote{Every point in the accretion disk is in local thermal equilibrium, and therefore we can define an effective temperature $T_{\rm eff}$ [in the case of an axisymmetric system, $T_{\rm eff} = T_{\rm eff} (r)$ depends only on the radial coordinate]. Different points have a different temperature, and therefore we speak about ``multi-color'' or ``multi-temperature'' spectrum. The gas temperature increases as the gas particles fall into the gravitational potential of the black hole and transform potential energy into kinetic and internal energy.}. For a given radius of the disk, the temperature depends on the black hole mass and the mass accretion rate. The peak temperature is reached near the inner edge of the disk and is in the soft X-ray band ($0.1-1$~keV) for stellar-mass black holes and in the optical/UV band ($1-10$~eV) for supermassive black holes.  The thermal component of the accretion disk is indicated by the red arrows in Fig.~\ref{f-corona}.

The {\it corona} is a hotter ($\sim 100$~keV), usually optically thin, cloud close to the black hole, but its exact geometry is currently unknown. In the {\it lamppost geometry}, the corona is a point-like source along the spin axis of the black hole~\cite{lamppost}. In the {\it sandwich geometry}, it is the atmosphere above the accretion disk~\cite{sandwich}. The inverse Compton scattering of the thermal photons from the accretion disk off free electrons in the corona produces a power-law component (blue arrows in Fig.~\ref{f-corona}) with a cut-off energy that depends on the temperature of the corona ($E_{\rm cut} \sim 100-1000$~keV).

The power-law component from the corona illuminates also the accretion disk, producing a reflection component (green arrows in Fig.~\ref{f-corona}) with some fluorescent emission lines~\cite{ref}. The strongest feature of the reflection component is usually the iron K$\alpha$ line, which is at 6.4~keV in the case of neutral or weakly ionized iron and shifts up to 6.97~keV for H-like iron ions.

\subsection{Spectral States \label{ss-ss}}

An accreting black hole can be found in different ``spectral states'', which are characterized by the luminosity of the source and by the relative contribution of its spectral components (thermal, power-law, reflection)~\cite{spst,spst2}. The spectral state classification is purely phenomenological, i.e. follows from the observed X-ray spectrum. However, there should be a correlation (not completely understood as of now) between spectral states and accretion flow configurations.

Let us start discussing the case of a stellar-mass black hole in an X-ray transient. The object typically spends most of the time in a {\it quiescent state} with a very low accretion luminosity ($L/L_{\rm Edd} < 10^{-6}$). At a certain point, the source has an {\it outburst} and becomes a bright X-ray source in the sky ($L/L_{\rm Edd} \sim 10^{-3} - 1$). The quiescent state is determined by a very low mass accretion rate, namely a very low amount of material transfers from the companion star to the black hole. When there is a sudden increase of the mass accretion rate (for instance, the companion star inflates and the black hole strips material from the surface of the companion), there is the outburst. The object may be in a quiescent state for several months or even decades. An outburst typically lasts from some days to a few months (roughly the time that the black hole takes to swallow the material that produced the outburst). During an outburst, the spectrum of the source changes.

\begin{figure}[b]
\begin{center}
\includegraphics[type=pdf,ext=.pdf,read=.pdf,width=8.7cm]{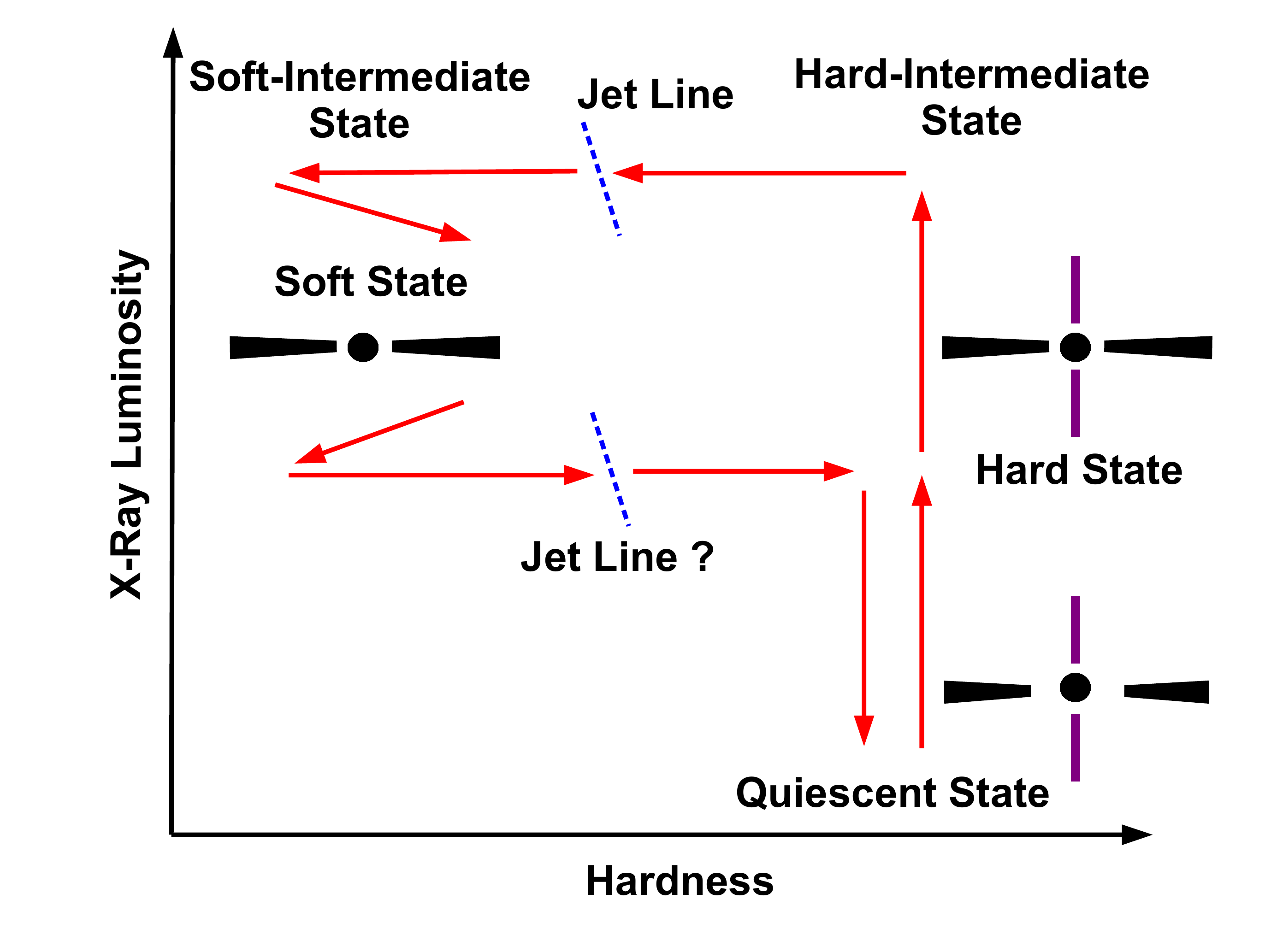}
\end{center}
\vspace{-0.5cm}
\caption{Evolution of the prototype of an outburst in the hardness-intensity diagram. The source is initially in a quiescent state. At the beginning of the outburst, the source enters the hard state, then moves to some intermediate states, to the soft state, and eventually returns to a quiescent state. See the text for more details. \label{f-hid}}
\end{figure}

The {\it hardness-intensity diagram} (HID)~\cite{spst,spst2} is a useful tool for the description of an outburst, see Fig.~\ref{f-hid}. The $x$-axis is for the hardness of the source, which is the ratio between its luminosity in the hard and soft X-ray bands, for instance between the luminosity in the $6-10$ and $2-6$~keV bands, but other choices are also common. The $y$-axis can be for the X-ray luminosity or the count number of the instrument, but other choices are also possible. The hardness-intensity diagram depends on the source (e.g. the interstellar absorption) and on the instrument (e.g. its effective area at different energies), but, despite that, it turns out to be very useful to study transient sources.

The relation between spectral states and accretion flow can be understood noting that the intensity of the thermal component is mainly determined by the mass accretion rate and the position of the inner edge of the accretion disk, while the contributions of the power-law and reflection components depend on the properties of the corona (its location, extension, geometry, etc.). In particular, the local flux of the disk's thermal component is approximately proportional to the mass accretion rate and the inverse of the cube of the disk's radius, $\mathcal{F}(r) \propto \dot{M}/r^3$~\cite{lasota4}. When the mass accretion rate is low (high) and the inner edge of the disk is at large (small) radii, the thermal component is weak (strong). The power-law and the reflection components are strong (weak) when the corona is large (small) and close to (far from) the disk. The relative contribution of these three components depends on the material around the black hole, and, in turn, determines the spectral state.

{\it Quiescent state ---} The source is initially in a quiescent state: the mass accretion rate and the luminosity are very low (the source may also be too faint to be detected) and the spectrum is hard. The inner edge of the accretion disk is truncated at a radius significantly larger than the radius of the ISCO.

{\it Hard state ---} At the beginning of the outburst, the spectrum is hard and the source becomes brighter and brighter because the mass accretion rate increases ($L/L_{\rm Edd}$ starts from $\sim 10^{-3}$ and can reach values up to $\sim 0.7$ in some cases). The spectrum is dominated by the power-law  and reflection components. The thermal component is subdominant, and the temperature of the inner part of the disk may be low, around 0.1~keV or even lower, but it increases as the luminosity of the source rises. The inner edge of the disk is initially at a radius larger than the ISCO one, but it moves to the ISCO as the luminosity increases (as shown in Fig.~\ref{f-hid}, where the disk is in black), and it may be at the ISCO at the end of the hard state. During the hard state, compact mildly relativistic steady jets (in violet in Fig.~\ref{f-hid}) are common, but the exact mechanism producing these jets is currently unknown.

{\it Intermediate states ---} 
The power-law and the reflection components get weaker, probably because of a variation in the geometry/properties of the corona. As a consequence, the contribution of the thermal component increases and the source moves to the left part of the HID. We first have the {\it hard-intermediate state} and then the {\it soft-intermediate state}. As shown in Fig.~\ref{f-hid}, there exists a {\it jet line}, not well understood for the moment, in the HID: when the source crosses the jet line, we observe transient highly relativistic jets. Even in this case, the mechanism responsible for the production of these jets is unknown. If the hardness of the source oscillates near the jet line, we can observe several transient jets.

{\it Soft state ---} The thermal spectrum of the disk is the dominant component in the spectrum and the inner part of the disk temperature is around 1~keV. If the luminosity of the source is between 5\% to 30\% of its Eddington luminosity, the inner edge of the disk should be at the ISCO radius~\cite{cfm-1915}. In the soft state, we do not observe any kind of jet\footnote{For instance, in the corona lamppost geometry, the corona may be the base of the jet. This could explain why, in the soft state, we do not see jets and the power-law and reflection components are weak.}. However, strong winds and outflows are common (while they are absent in the hard state). The luminosity of the source may somewhat decreases and changes hardness, remaining on the left side of the HID.

At a certain point, the transfer of material decreases, leading to the end of the outburst. The contribution of the thermal spectrum of the disk decreases and, as a consequence, the hardness of the source increases. The source re-enters the soft-intermediate state, the hard-intermediate state, then the hard state, and eventually, when the hardness is high, the luminosity drops down and the source returns to the quiescent state till the next outburst. Between the soft-intermediate and the hard-intermediate states, we may observe transient jets, but the existence of a jet line is not clear here. Every source follows the path shown in Fig.~\ref{f-hid} counter-clockwise, but there are differences among different sources and even for the same source among different outbursts.

In the case of stellar-mass black holes in persistent X-ray sources, there is no outburst, but we can still use the HID. The most studied source is Cygnus~X-1 (the other persistent sources are in nearby galaxies, so they are fainter and more difficult to study). This object spends most of the time in the hard state, but it occasionally moves to a softer state, which is usually interpreted as a soft state. LMC X-1 is always in the soft state. LMC X-3 is usually observed in the soft state, rarely in the hard state, and there is no clear evidence that this source can be in an intermediate state.

In the case of supermassive black holes, there are at least two important differences. First, the size of the system, which scales as the mass. 1~day for a 10~$M_\odot$ black hole corresponds to 3,000~years for a $10^7$~$M_\odot$ black hole, which makes impossible the study of the evolution of a specific system. Second, the temperature of the disk is in the optical/UV range for a supermassive black hole. Despite these two issues, stellar-mass and supermassive black holes have a similar behavior and we can employ the same spectral state classification (see, for instance, \cite{spst} and references therein).


\section{Techniques for Probing the Strong Gravity Region}

The aim of this section is to describe the leading techniques to probe the strong gravity region around black holes. The continuum-fitting method and X-ray reflection spectroscopy are well-established electromagnetic approaches. The measurement of quasi-periodic oscillations is not yet a mature technique, because several models have been proposed but we do not know which one, if any, is correct. Direct imaging of the accretion flow will be possible very soon for SgrA$^*$, the supermassive black hole at the center of our Galaxy. Gravitational waves are a recent new tool that promises a huge amount of completely new data in the next years.

\subsection{Continuum-Fitting Method}

Within the Novikov-Thorne model~\cite{ntm,ntm2}, we can derive the time-averaged radial structure of the accretion disk from the fundamental laws of the conservation of rest-mass, energy, and angular momentum. The time-averaged energy flux emitted from the surface of the disk is
\be
\mathcal{F} (r) = \frac{\dot{M} c^2}{4 \pi r_{\rm g}^2} F(r) \, ,
\ee
where $\dot{M}=dM/dt$ is the time-averaged mass accretion rate, which is independent of the radial coordinate, and $F(r)$ is a dimensionless function of the radial coordinate that becomes roughly of order 1 at the disk inner edge (see~\cite{ntm2} for more details). Assuming that the disk is in local thermal equilibrium, its emission is blackbody-like and at any radius we can define an effective temperature $T_{\rm eff} (r)$ from the time-averaged energy flux as $\mathcal{F} = \sigma T^4_{\rm eff}$, where $\sigma$ is the Stefan-Boltzmann constant.

Novikov-Thorne disks with the inner edge at the ISCO radius are realized when the accretion luminosity is between 5\% to 30\% of the Eddington limit of the object~\cite{cfm-1915}, and this is confirmed by theoretical~\cite{edd1a,edd1b} and observational studies~\cite{edd2}. At lower luminosities, the disk is more likely truncated. At higher luminosities, the gas pressure becomes important, the inner part of the disk is not thin any longer, and the inner edge might be at a radius smaller than the ISCO. Requiring $\dot{M} \sim 0.1 \, \dot{M}_{\rm Edd}$ as the condition for Novikov-Thorne disks, we can get a rough estimate of the effective temperature of the inner part of the accretion disk
\be
T_{\rm eff} \sim 
\left( \frac{0.1 \; \dot{M}_{\rm Edd} c^2}{4 \pi \sigma r_{\rm g}^2} \right)^{1/4} 
\sim \left( \frac{10 \; M_\odot}{M} \right)^{1/4} \text{keV} \, ,
\ee
and we can see that the disk's thermal spectrum is in the soft X-ray band for stellar-mass black holes and in the optical/UV band for the supermassive one.

The continuum-fitting method is the analysis of the thermal spectrum of geometrically thin and optically thick accretion disks of black holes in order to measure the black hole spin parameter $a_*$~\cite{cfm1,cfm2,cfm3,cfm4}. The technique is normally used for stellar-mass black holes only, because the spectrum of supermassive black holes is in the optical/UV band where dust absorption limits the capability of accurate measurements.

The model describing the thermal spectrum of an accretion disk around a Kerr black holes depends on five parameters: the black hole mass $M$, the mass accretion rate $\dot{M}$, the inclination angle of the disk $i$, the distance of the source from the observer $D$, and the black hole spin parameter $a_*$. It is not possible to infer all these parameters from the data of the spectrum of a thin disk, because there is a degeneracy. However, if we can get independent measurements of $M$, $D$, and $i$, usually from optical observations, it is possible to fit the thermal component and measure $a_*$ and $\dot{M}$. This is the continuum-fitting method. Currently, there are about ten stellar-mass black holes with a spin measurement from the continuum-fitting method, see Tab.~\ref{t-spin}.

\begin{table*}[t]
\centering
\begin{tabular}{|ccccccc|}
\hline 
BH Binary & \hspace{0.1cm} & $a_*$ (Continuum) & \hspace{0.1cm} & $a_*$ (Iron) & \hspace{0.1cm} & Principal References \\
\hline 
GRS~1915+105 && $> 0.98$ && $0.98 \pm 0.01$ && \cite{cfm-1915,cfm-1915b} \\
Cyg~X-1 && $> 0.98$ && $0.97^{+0.014}_{-0.02}$ && \cite{cfm-cyg1,cfm-cyg2,cfm-cyg3,cfm-cyg4,cfm-cyg5,cfm-cyg6} \\
GS 1354-645 && -- && $> 0.98$ && \cite{cfm-gs1354} \\
LMC~X-1 && $0.92 \pm 0.06$ && $0.97^{+0.02}_{-0.25}$ && \cite{cfm-lmcx1,cfm-lmcx1b} \\
GX~339-4 && $< 0.9$ && $0.95\pm0.03$ && \cite{cfm-gx339,cfm-gx339b,cfm-gx339c,cfm-gx339d} \\
MAXI~J1836-194 && --- && $0.88 \pm 0.03$ && \cite{cfm-maxi} \\
M33~X-7 && $0.84 \pm 0.05$ && --- && \cite{cfm-liu08} \\
4U~1543-47 && $0.80 \pm 0.10^\star$ && --- && \cite{cfm-sh06} \\
IC10 X-1     &&  $\gtrsim0.7$  && --- && \cite{cfm-st16} \\
Swift~J1753.5 && --- && $0.76^{+0.11}_{-0.15}$ && \cite{cfm-swift} \\
XTE~J1650-500 && --- && $0.84 \sim 0.98$ && \cite{cfm-1650} \\
GRO~J1655-40 && $0.70 \pm 0.10^\star$ && $> 0.9$ && \cite{cfm-sh06,cfm-swift} \\
GS~1124-683 && $0.63^{+0.16}_{-0.19}$ && --- && \cite{cfm-gou_novamus} \\
XTE~J1752-223 && --- && $0.52 \pm 0.11$ && \cite{cfm-1752} \\
XTE~J1652-453 && --- && $< 0.5$ && \cite{cfm-1652} \\
XTE~J1550-564 && $0.34 \pm 0.28$ && $0.55^{+0.15}_{-0.22}$ && \cite{cfm-xte} \\
LMC~X-3 && $0.25 \pm 0.15$ && --- && \cite{cfm-lmcx3} \\
H1743-322 && $0.2 \pm 0.3$ && --- && \cite{cfm-h1743} \\
A0620-00 &&  $0.12 \pm 0.19$ && --- && \cite{cfm-62} \\
XMMU~J004243.6 && $< -0.2$ && --- && \cite{cfm-m31} \\
\hline 
\end{tabular}
\vspace{0.4cm}
\caption{Summary of the continuum-fitting and iron line measurements of the spin parameter of stellar-mass black holes. See the references in the last column for more details. Note: $^\star$These sources were studied in an early work of the continuum-fitting method, within a more simple model, and therefore the published 1-$\sigma$ error estimates are doubled following~\cite{cfm4}. \label{t-spin}}
\end{table*}

\begin{table}[t]
\centering
\begin{tabular}{|ccccc|}
\hline 
Object & \hspace{0.1cm} & $a_*$ (Iron) & \hspace{0.1cm} & Principal References \\
\hline 
IRAS13224-3809 && $> 0.99$ && \cite{suzaku} \\
Mrk110 && $> 0.99$ && \cite{suzaku} \\
NGC4051 && $> 0.99$ && \cite{ngc4051} \\
1H0707-495 && $> 0.98$ && \cite{suzaku,1h0707} \\
RBS1124 && $> 0.98$ && \cite{suzaku} \\
NGC3783 && $> 0.98$ && \cite{ngc3783} \\
NGC1365 && $0.97^{+0.01}_{-0.04}$ && \cite{ngc1365a,ngc1365b} \\
Swift~J0501-3239 && $> 0.96$ && \cite{suzaku} \\
PDS456 && $> 0.96$ && \cite{suzaku} \\
Ark564 && $0.96^{+0.01}_{-0.06}$ && \cite{suzaku} \\
3C120 && $> 0.95$ && \cite{3c120} \\
Mrk79 && $> 0.95$ && \cite{mrk79} \\
MCG-6-30-15 && $0.91^{+0.06}_{-0.07}$ && \cite{mcg63015a,mcg63015b} \\
TonS180 && $0.91^{+0.02}_{-0.09}$ && \cite{suzaku} \\
1H0419-577 && $> 0.88$ && \cite{suzaku} \\
IRAS00521-7054 && $> 0.84$ && \cite{iras521} \\
Mrk335 && $0.83^{+0.10}_{-0.13}$ && \cite{suzaku,mrk335} \\
Ark120 && $0.81^{+0.10}_{-0.18}$ && \cite{suzaku,ark120} \\
Swift~J2127+5654 && $0.6^{+0.2}_{-0.2}$ && \cite{swift2127} \\
Mrk841 && $> 0.56$ && \cite{suzaku} \\
Fairall9 && $0.52^{+0.19}_{-0.15}$ && \cite{suzaku,fairall9} \\
\hline 
\end{tabular}
\vspace{0.4cm}
\caption{Summary of spin measurements of supermassive black holes reported in the literature. See the references in the last column for more details. \label{t-spin-agn}}
\end{table}

\subsection{X-Ray Reflection Spectroscopy}

X-ray reflection spectroscopy (or iron line method) refers to the study of the reflection component. This technique can be applied to both stellar-mass and supermassive black holes and is currently the only available method to measure the spin of supermassive black holes~\cite{i1,i2}.

The most prominent feature of the reflection spectrum is usually the iron K$\alpha$ line. This is because the iron is more abundant than other heavy elements (the iron-26 nucleus is more tightly bound than lighter and heavier elements, so it is the final product of nuclear reactions) and the probability of fluorescent line emission is also high. The iron K$\alpha$ line is a very narrow feature in the rest-frame of the emitter, while the one observed in the reflection spectrum of black holes can be very broad and skewed, as the result of relativistic effects occurring in the strong gravity region of the object (gravitational redshift, Doppler boosting, light bending)~\cite{book,i1,i2,i2003}. While the iron K$\alpha$ line is usually the strongest feature, accurate measurements of black hole spins require to fit the whole reflection spectrum, not just the iron line.

Reflection models describing the reflection component of accretion disks around Kerr black holes depend on several parameters: the black holes spin $a_*$, the inner edge of the disk $R_{\rm in}$ (which may or may not be at the ISCO radius, see the discussion in Section~\ref{ss-ss}), the outer edge of the disk $R_{\rm out}$, the inclination angle of the disk $i$, the iron abundance $A_{\rm Fe}$, the ionization of the disk $\xi$, and some parameters related to the emissivity profile of the disk. The latter is quite a crucial ingredient and depends on the geometry of the corona, which is currently unknown. Coronas with arbitrary geometries can be modeled with a power-law emissivity profile (the intensity on the disk is $I \propto 1/r^q$ where $q$ is the emissivity index) or with a broken power-law ($I \propto 1/r^{q_{\rm in}}$ for $r < R_{\rm br}$, $I \propto 1/r^{q_{\rm out}}$ for $r > R_{\rm br}$, and we have three parameters: the inner emissivity index $q_{\rm in}$, the outer emissivity index $q_{\rm out}$, and the breaking radius $R_{\rm br}$). In the case of supermassive black holes, it is often necessary to take the cosmological redshift $z$ into account. For stellar-mass black holes, their relative motion in the Galaxy is of order 100~km/s and their Doppler boosting can be ignored.

Note that spin measurements with the iron line method do not require independent measurements of the black hole mass $M$, the distance $D$, and the inclination angle of the disk $i$, three quantities that are required in the continuum-fitting method, are usually difficult to measure, and have large uncertainty. The reflection spectrum is independent of $M$ and $D$, and can directly measure the inclination angle of the disk $i$.

Current spin measurements of stellar-mass black holes with the iron line method are summarized in the third column in Tab.~\ref{t-spin} (see the corresponding references in the fourth column for more details). Note that some black holes have their spin measured with both the continuum-fitting and the iron line methods. In general, the two measurements agree (GRS~1915+105, Cyg~X-1, LMC~X-1, XTE~J1550-564). For GX~339-4 and GRO~J1655-40, the two measurements are not consistent. The iron line method is usually applied when the source is in the hard state, when the reflection spectrum is stronger but the disk may be truncated at a radius larger than the ISCO. This would lead to underestimate the black hole spin, and therefore it cannot be the case of the spin measurements of GX~339-4 and GRO~J1655-40, where the iron line method provides spin values higher than the continuum-fitting method. As pointed out before, the continuum-fitting method crucially depends on independent measurements of the black hole mass $M$, the distance $D$, and the inclination angle of the disk $i$, three quantities that are usually difficult to measure and may be affected by systematic effects. For example, in the case of GRO~J1655-40 there are a few mass measurements reported in the literature, but they are not consistent among them.

A summary of spin measurements of supermassive black holes with the iron line method is reported in Tab.~\ref{t-spin-agn} (see the references in the last column for more details and the lists of spin measurements in~\cite{i1,i2,vasudevan} for a few more sources with a constrained spin). Note the very high spin of several objects. In part, this can be explained noting that fast-rotating black holes are brighter and thus the spin measurement is easier. If these measurements are correct, they would point out that these objects have been spun up by prolonged disk accretion and therefore would provide information about galaxy evolutions (see the discussion in Section~\ref{ss-evolution}). However, the very high spin measurements have to be taken with some caution, as they may be affected by systematic effects in the model employed to infer the black hole spin. More details on the possible interpretation of current spin measurements of supermassive black holes can be found in~\cite{i1}.

\subsection{Quasi-Periodic Oscillations \label{ss-qpos}}

\begin{figure}[b]
\vspace{-0.3cm}
\begin{center}
\includegraphics[type=pdf,ext=.pdf,read=.pdf,width=8.7cm]{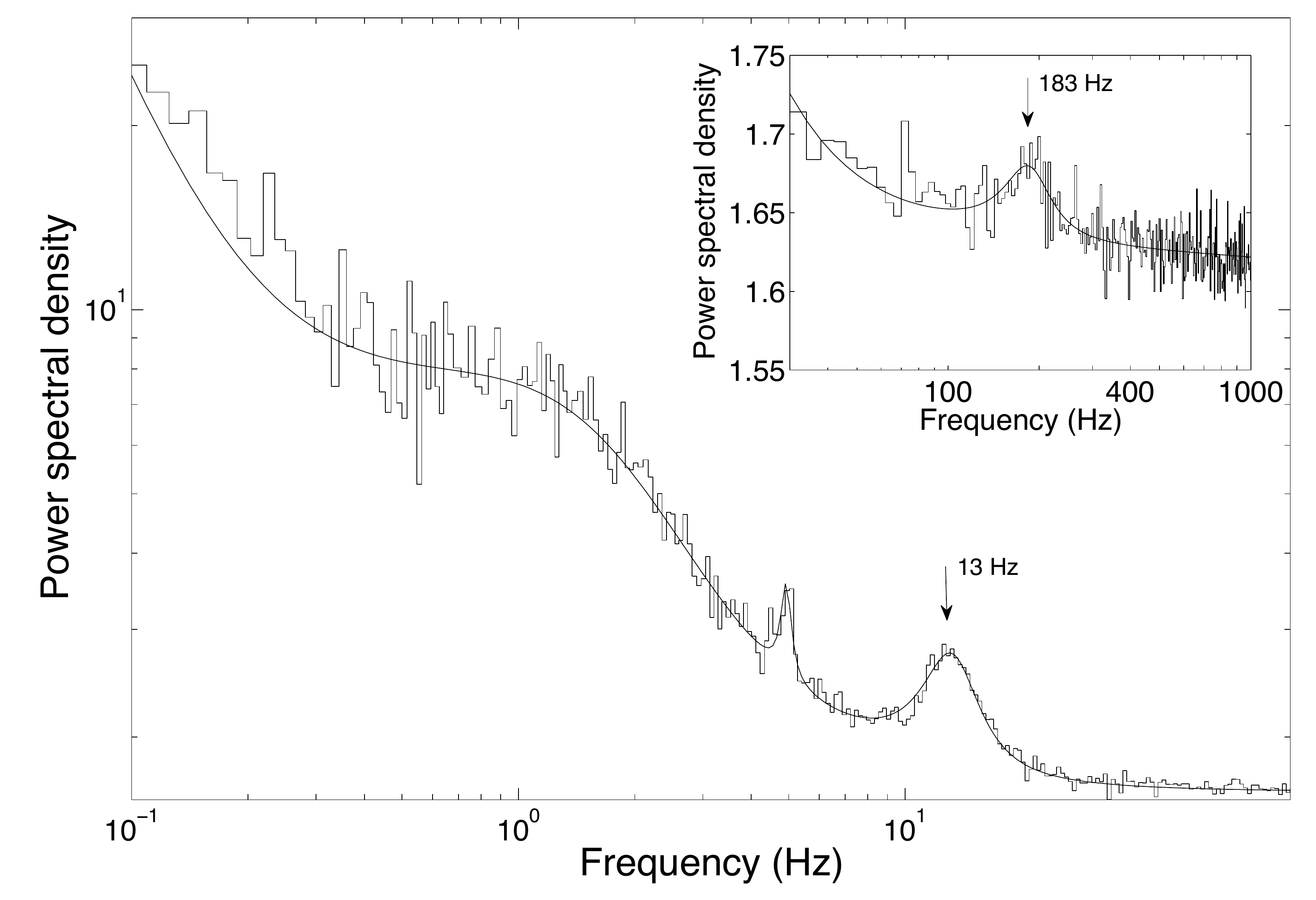}
\end{center}
\vspace{-0.5cm}
\caption{Power density spectrum from an observation of XTE~J1550-564. We see a QPO around 5~Hz, a QPO at 13~Hz (marked by an arrow), and a QPO at 183~Hz in the inset (marked by an arrow). Fig.~1 from~\cite{motta14}, reproduced by permission of Oxford University Press. \label{f-qpos}}
\end{figure}

Quasi-periodic oscillations (QPOs) are a common feature in the X-ray power density spectrum of neutron stars and stellar-mass black holes~\cite{vdk}. The power density spectrum $P(\nu)$ is the square of the Fourier transform of the photon count rate $C(t)$. If we use the Leahy normalization, we have
\be
P ( \nu ) = \frac{2}{N} 
\left| \int_0^T C(t) e^{-2 \pi i \nu t} dt \right| \, ,
\ee
where $N$ is the total number of counts and $T$ is the duration of the observation. QPOs are narrow features in the X-ray power density spectrum of a source. Fig.~\ref{f-qpos} shows the power density spectrum obtained from an observation of the stellar-mass black hole XTE~J11550-564, where we can see a QPO around 5~Hz, one at 13~Hz, and one at 183~Hz in the inset.

In the case of black hole binaries, QPOs can be grouped into two classes: low-frequency QPOs ($0.1-30$~Hz) and high-frequency QPOs ($40-450$~Hz). The exact nature of these QPOs is currently unknown, but there are several proposals in literature. In most scenarios, the frequencies of the QPOs is somehow related to the fundamental frequencies of a particle orbiting the black hole~\cite{qqq1,qqq2,qqq3}:
\begin{enumerate}
\item {\it Orbital frequency} $\nu_\phi$, which is the inverse of the orbital period.
\item {\it Radial epicyclic frequency} $\nu_r$, which is the frequency of radial oscillations around the mean orbit.
\item {\it Vertical epicyclic frequency} $\nu_\theta$, which is the frequency of vertical oscillations around the mean orbit.
\end{enumerate}

In the Kerr metric, we have a compact analytic form for the expression of these frequencies
\be
\hspace{-0.4cm}
\nu_\phi &=& \frac{c}{2\pi} \sqrt{\frac{r_{\rm g}}{r^3}}
\left[1 \pm a_* \left(\frac{r_{\rm g}}{r}\right)^{3/2} \right]^{-1} \, , \\
\hspace{-0.4cm}
\nu_r &=& \nu_\phi \sqrt{1 - 6 \, \frac{r_{\rm g}}{r} 
\pm 8 a_* \left(\frac{r_{\rm g}}{r}\right)^{3/2} 
- 3 a^2_* \left(\frac{r_{\rm g}}{r}\right)^2} \, , \\
\hspace{-0.4cm}
\nu_\theta &=& \nu_\phi \sqrt{1 \mp 4 a_* \left(\frac{r_{\rm g}}{r}\right)^{3/2} 
+ 3 a^2_* \left(\frac{r_{\rm g}}{r}\right)^2} \, ,
\ee
where $r$ is the orbital radius in Boyer-Lindquist coordinates.

To have an idea of the order of magnitude of these frequencies, we can write the orbital frequency for a Schwarzschild black hole
\be
\nu_\phi (a_* = 0) = 220 \left(\frac{10 \, M_\odot}{M}\right) 
\left(\frac{6 \, r_{\rm g}}{r}\right)^{3/2} \text{Hz} \, .
\ee
High-frequency QPOs at $40-450$~Hz are thus of the right magnitude to be associated to the orbital frequencies near the ISCO radius of stellar-mass black holes. Interestingly, we have evidence also of high-frequency QPOs in supermassive black holes ($< 1$~mHz)~\cite{qpo-smbh} and intermediate-mass black holes ($\sim 1$~Hz)~\cite{qpo-imbh}.

\subsection{Direct Imaging}

\begin{figure}[t]
\begin{center}
\includegraphics[type=pdf,ext=.pdf,read=.pdf,width=8.7cm]{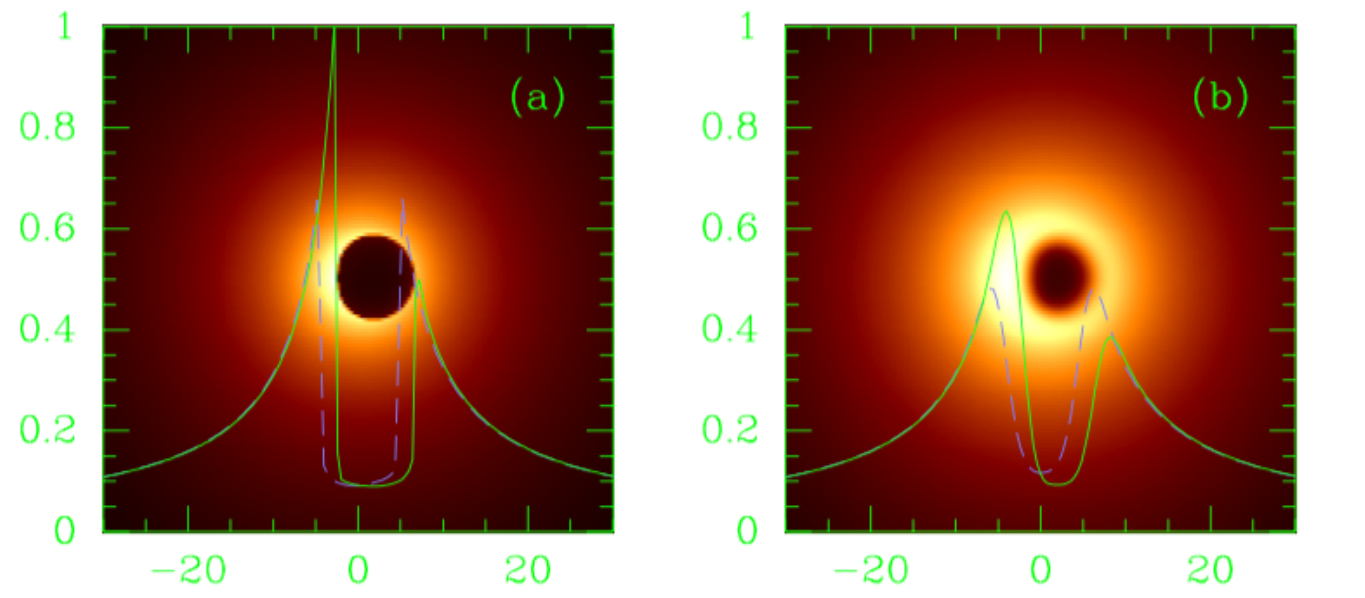}
\end{center}
\vspace{-0.2cm}
\caption{Direct image of a black hole surrounded by an optically thin emitting medium with the characteristics of that of SgrA$^*$. The black hole spin parameter is $a_* = 0.998$ and the viewing angle is $i=45^\circ$. Panel~$a$: image from ray-tracing calculations. Panel~$b$: image from a simulated observation of an idealized VLBI experiment at 0.6~mm wavelength taking interstellar scattering into account. The solid green curve and the dashed purple curve show, respectively, the intensity variations of the image along the $x$-axis and the $y$-axis. From~\cite{shadow}. \copyright AAS. Reproduced with permission. \label{f-shadow}}
\end{figure}

Depending on the geometry of the accretion disk and on its optically properties (thin/thick), if we could image the accretion flow around a black hole with a resolution of at least some gravitational radii, we would observe a dark area of a brighter background. The dark area is usually referred to as the black hole {\it shadow}~\cite{shadow} (see Fig.~\ref{f-shadow}) and its boundary is determined by light bending in the strong gravity region~\cite{bardeen73}.

Very long baseline interferometry (VLBI) facilities are the combination of several radio telescopes far each other that are used as a single telescope to reach much smaller angular resolutions. Sub-mm VLBI experiments promise to observe the shadow of the supermassive black hole at the center of the Galaxy very soon~\cite{eht}. Its mass is about $4 \cdot 10^6$~$M_\odot$ and it is at $d \approx 8$~kpc from us, so its angular size in the sky is roughly
\be
\theta \sim \frac{r_{\rm g}}{d} \sim 0.05 \text{ milliarcseconds} \, .
\ee

There are three particular conditions that make the observation of the shadow of the black hole of the Galaxy achievable. $i)$ The angular resolution of VLBI facilities scales as $\lambda/D$, where $\lambda$ is the electromagnetic radiation wavelength and $D$ is the distance among different stations. For $\lambda < 1$~mm and stations located in different continents ($D > 10^3$~km), it is possible to reach an angular resolution of 0.1~milliarcseconds. $ii)$ The emitting medium around the black hole at the center of the Galaxy is optically thick at wavelength $\lambda > 1$~mm, but becomes optically thin for $\lambda < 1$~mm. $iii)$ The interstellar scattering at the center of the Galaxy dominates over intrinsic source structures at wavelength $\lambda > 1$~mm, but becomes subdominant for $\lambda < 1$~mm.

In the case of stellar-mass black holes in the Galaxy, the angular size is 4-5 orders of magnitude smaller. Similar angular resolutions are impossible today, but they may be possible in the future with X-ray interferometric techniques~\cite{maxim}.

\subsection{Gravitational Waves \label{ss-gw}}

Gravitational waves are produced by the motion of massive bodies like electromagnetic waves are produced by the motion of charged particles (for an introductory review on gravitational waves and gravitational wave astronomy, see, for instance, \cite{ggww1,ggww2,ggww3}). In the case of black holes, we expect the emission of gravitational waves when they are in a binary system. The motion of the two bodies generates gravitational waves, the system loses energy and angular momentum, the orbital separation and the orbital period decrease, and eventually the two bodies merge together to create a larger black hole~\cite{ggww2}.

Depending on the size of the system, the frequency of the gravitational wave signal changes. In the case of stellar-mass objects (a binary system with two stellar-mass black holes or a system of a stellar-mass black hole and a neutron star), we can observe the signal for fractions of a second/minute before the merger with ground-based laser interferometers (like LIGO~\cite{ligoexp} or Virgo~\cite{virgo}), which work in the frequency range 10~Hz to 10~kHz. In the case of binary systems of two black holes with masses $M \sim 10^6$~$M_\odot$, as well as in the case of a stellar-mass compact object (black hole, neutron star or white dwarf) orbiting a black hole of mass $M \sim 10^6$~$M_\odot$, the frequency of the gravitational waves is in the range $0.1-100$~mHz and can be detected by future space-based laser interferometers (like LISA~\cite{elisa,elisa2,elisa3} or DECIGO~\cite{decigo}). For binary systems with black holes of $10^9$~$M_\odot$, the gravitational wave frequencies are in the range $1-100$~nHz and the signal can be detected by pulsar timing array experiments~\cite{pulsar}.

The first gravitational wave event was detected on 14~September 2015 by the LIGO experiment and called GW150914~\cite{gw150914}. LIGO consists of two ground-based laser interferometers, one in the state of Washington and one in Louisiana. GW150914 was produced by the coalescence of two black holes, each of them with a mass around 30~$M_\odot$, and the final product was a black hole of about 60~$M_\odot$. Other events were detected in~2017 (there was no detection in~2016 because the facilities were under upgrading). The detection of many other events by the LIGO/Virgo collaboration is expected in the next years, but these facilities can only detect gravitational waves from stellar-mass black holes because for larger masses the gravitational wave frequency is too low.

\begin{figure}[t]
\begin{center}
\includegraphics[type=pdf,ext=.pdf,read=.pdf,width=8.7cm]{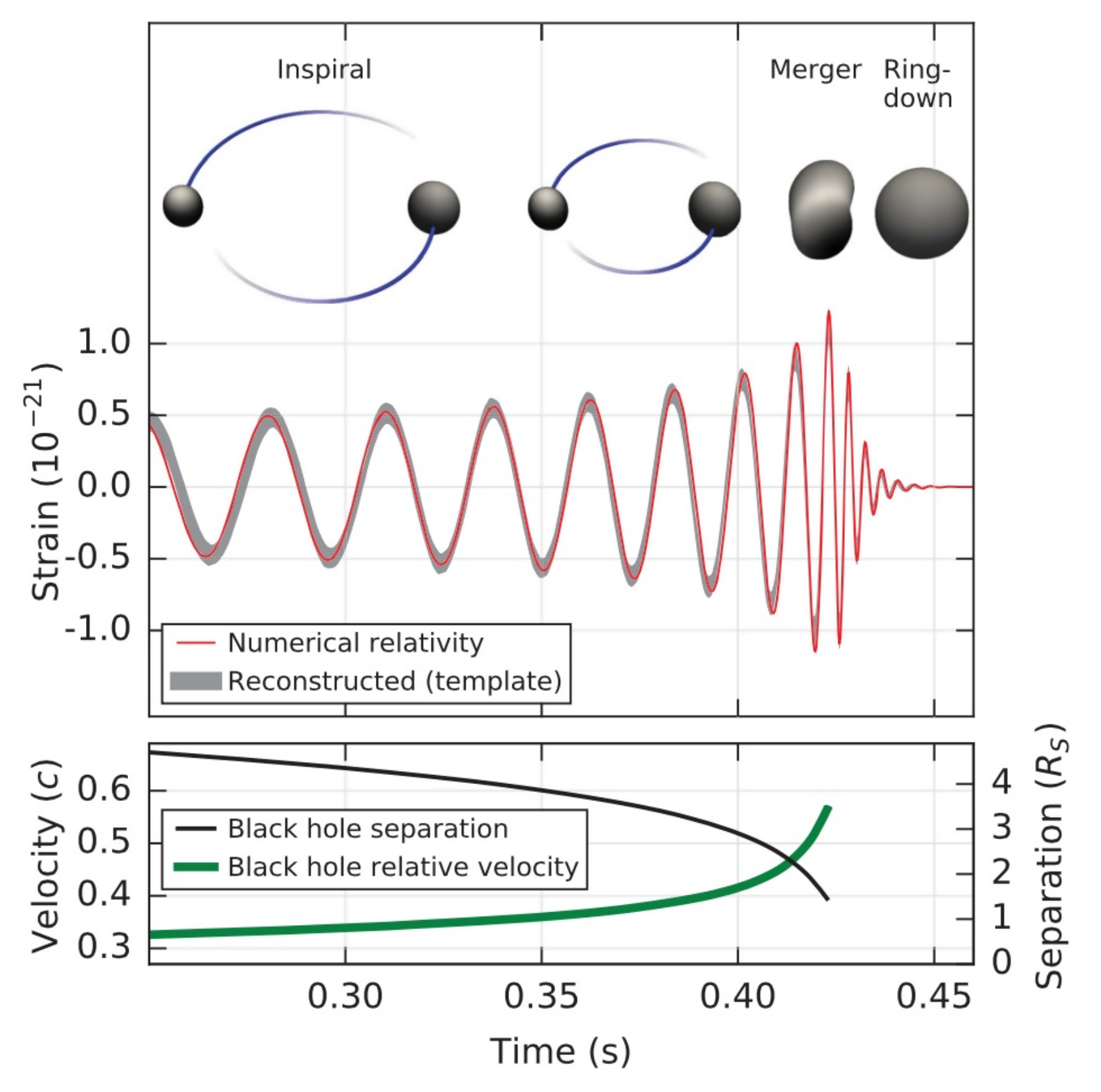}
\end{center}
\vspace{-0.2cm}
\caption{Temporal evolution of the strain, the black hole separation, and the black hole relative velocity in the event GW150914. From~\cite{gw150914} under the terms of the Creative Commons Attribution 3.0 License. \label{f-gw150914}}
\end{figure}

The coalescence of a binary black hole is characterized by three stages: {\it inspiral}, {\it merger}, and {\it ring-down} (see Fig.~\ref{f-gw150914}).

During the inspiral phase, the two objects orbit around each other emitting gravitational waves. The calculations of the evolution of the system and of the gravitational wave signal are usually based on post-Newtonian methods, where the expansion parameter is $\epsilon \sim U \sim v^2$, $U$ is the Newtonian potential, and $v$ is the black hole relative velocity~\cite{pnm}. The {\it chirp mass} is $\mathcal{M} = M^{2/5} \mu^{3/5}$, where $M = M_1 + M_2$, $\mu = M_1 M_2 / M$, and $M_1$ and $M_2$ are the masses of the two black holes. The chirp mass of a binary black hole emitting gravitational waves can be inferred by measuring the observed wave frequency $f$ and its time derivative $\dot{f}$~\cite{pnm}\footnote{Note that $f$ is the observed frequency, i.e. the frequency measured in the detector frame. For cosmological sources, $f$ does not coincide with the emission frequency in the source frame. If we use the observed $f$ and $\dot{f}$ in the right hand side in Eq.~(\ref{eq-chirp}), the left hand side should be the {\it redshifted chirp mass} $\mathcal{M}' = \mathcal{M} (1 + z)$, where $z$ is the cosmological redshift of the source.} 
\be\label{eq-chirp}
\mathcal{M} = \frac{c^3}{G_{\rm N}} \left( \frac{5}{96 \pi^{8/3}} \frac{\dot{f}}{f^{11/3}}\right)^{3/5} \, .
\ee

The passage of a gravitational wave is detected by a gravitational wave detector as a variation $\Delta L$ of some reference distance $L$. The quantity measured by the gravitational wave detector is the {\it strain} $h = \Delta L/L$. For a binary black hole in the Newtonian circular binary approximation, the amplitude of the strain is
\be
h = \frac{4 \pi^{2/3} G_{\rm N}^{5/3}}{c^4} \frac{\mathcal{M}^{5/3} f^{2/3}}{r} \, ,
\ee
where $r$ is the distance of the detector from the source. The waveform peaks at the merger (as shown in Fig.~\ref{f-gw150914}). A rough estimate of the peak strain can be obtained by using the gravitational wave frequency $f$ at the ISCO just before merger
\be
f = 2 f_{\rm orb} = \frac{1}{\pi} \sqrt{\frac{G_{\rm N} M}{r^3_{\rm ISCO}}} 
= \frac{1}{\pi} \frac{c^3}{\kappa \, G_{\rm N} M} \, ,
\ee
where $f_{\rm orb}$ is the orbital frequency and $\kappa = (r_{\rm ISCO}/r_{\rm g})^{3/2}$ is a numerical factor larger than 1. Eventually we find that $h_{\rm peak} \sim 0.1 \, ( r_{\rm g}/r )$ (this is a rough estimate assuming approximately equal mass black holes, so that $M$, $\mathcal{M}$, and $\mu$ are of the same order). Note that $h \propto 1/r$ has an important observational implication: if we improve the detector sensitivity by a factor 2, we increase the monitored volume, and therefore the detection rate, by a factor 8.

In the merger, the two black holes merge into a single black hole. When the system approaches the merger, the Post-Newtonian methods break down, because $\epsilon$ is not a small parameter any longer, and the description of the system requires to solve the Einstein equations numerically~\cite{numrel0,numrel1,numrel2,numrel3}. For non-spinning black holes, the stage of merger smoothly connects the stages of inspiral and ring-down. For rotating black holes, this stage may be a more violent event, depending on the black hole spins and their alignments with respect to the orbital angular momentum.

Lastly, the newly-born black hole emits gravitational waves to settle down to an equilibrium configuration. This is the ring-down stage, which is commonly described by black hole perturbation theory~\cite{teu,chandra}. The gravitational wave signal is characterized by damped oscillations, the so-called {\it quasi-normal modes}~\cite{kkk1,kkk2,bcw}. This is not a special property of black holes, and it is expected even in neutron stars or other possible compact self-gravitating systems (e.g. boson stars, wormholes, etc). However, the frequency and the damping time of these oscillations do depend on the specific system and its properties, while they are independent of the initial perturbations. In the case of a Schwarzschild black hole, the spectrum of the quasi-normal modes is only determined by the black hole mass. The frequency and the damping time of the dominant quasi-normal mode are~\cite{bcw}
\be
f &=& 1.207 \left( \frac{10 \, M_\odot}{M} \right) {\rm kHz} \, , \nonumber\\
\tau &=& 0.554 \left( \frac{M}{10 \, M_\odot} \right) {\rm ms} \, .
\ee
For a Kerr black hole, the spectrum depends on the mass $M$ and the spin parameter $a_*$.


\section{Concluding Remarks}

This article has briefly reviewed the state of the art of the study of astrophysical black holes, starting from their key-properties predicted in general relativity and summarizing current observations and techniques to probe the strong gravitational field of these objects. The references provided in each section should serve as the next step for the reader interested to know more on specific topics.

The past 10-20~years have seen significant progresses in the understanding of these systems, and we can expect much more progresses in the near future, thanks to the detection of a large number of gravitational wave events, the next generation of X-ray facilities (like eXTP~\cite{extp,snzhang} and Athena~\cite{athena}), and the observation of the shadow of the black hole at the center of our Galaxy with sub-mm VLBI experiments. While in the past this research field was studied only by astronomers, there is now an increasing interest from the physics community, because observational data are now reaching a level that it is possible to test fundamental physics, in particular general relativity and alternative theories of gravity~\cite{review,zheng}. There are also arguments suggesting that astrophysical black holes may be macroscopic quantum objects, and that we might be able to detect signatures of quantum gravity from their observations~\cite{quantum}.


\begin{acknowledgments}
The author acknowledges support from the National Natural Science Foundation of China (Grant No.~U1531117), Fudan University (Grant No.~IDH1512060), and the Alexander von Humboldt Foundation.
\end{acknowledgments}



\begin{thebibliography}{99}
  
\bibitem{book} 
  C.~Bambi,
  {\it Black Holes: A Laboratory for Testing Strong Gravity}
  (Springer Singapore, 2017),  doi:10.1007/978-981-10-4524-0
  
\bibitem{ein} 
  A.~Einstein,
  Annalen Phys.\  {\bf 49}, 769 (1916)
  [Annalen Phys.\  {\bf 14}, 517 (2005)].  
  
\bibitem{sch} 
  K.~Schwarzschild,
  Sitzungsber.\ Preuss.\ Akad.\ Wiss.\ Berlin (Math.\ Phys.\ ) {\bf 1916}, 189 (1916)
  [physics/9905030].   
  
\bibitem{fink} 
  D.~Finkelstein,
  Phys.\ Rev.\  {\bf 110}, 965 (1958).  
  
\bibitem{kerr} 
  R.~P.~Kerr,
  Phys.\ Rev.\ Lett.\  {\bf 11}, 237 (1963).  
  
\bibitem{zeld} 
  Y.~B.~Zeldovich,
  Dokl.\ Akad.\ Nauk\  {\bf 155}, 67 (1964)
  [Sov.\ Phys.\ Dokl.\  {\bf 9}, 195 (1964)].  
  
\bibitem{salp} 
  E.~E.~Salpeter,
  Astrophys.\ J.\  {\bf 140}, 796 (1964).  
  
\bibitem{cyg2} 
  C.~T.~Bolton,
  Nature {\bf 235}, 271 (1972).  
  
\bibitem{cyg3} 
  B.~L.~Webster and P.~Murdin,
  Nature {\bf 235}, 37 (1972).  
  
\bibitem{re-mc} 
  R.~A.~Remillard and J.~E.~McClintock,
  Ann.\ Rev.\ Astron.\ Astrophys.\  {\bf 44}, 49 (2006)
  [astro-ph/0606352].  
  
\bibitem{k-r} 
  J.~Kormendy and D.~Richstone,
  Ann.\ Rev.\ Astron.\ Astrophys.\  {\bf 33}, 581 (1995).     
  
\bibitem{gw150914} 
  B.~P.~Abbott {\it et al.} [LIGO Scientific and Virgo Collaborations],
  Phys.\ Rev.\ Lett.\  {\bf 116}, 061102 (2016)
  [arXiv:1602.03837 [gr-qc]].  
  
\bibitem{h1} 
  W.~Israel,
  Phys.\ Rev.\  {\bf 164}, 1776 (1967).

\bibitem{h2} 
  B.~Carter,
  Phys.\ Rev.\ Lett.\  {\bf 26}, 331 (1971).

\bibitem{h3} 
  D.~C.~Robinson,
  Phys.\ Rev.\ Lett.\  {\bf 34}, 905 (1975).

\bibitem{h4} 
  P.~T.~Chrusciel, J.~L.~Costa and M.~Heusler,
  Living Rev.\ Rel.\  {\bf 15}, 7 (2012)
  [arXiv:1205.6112 [gr-qc]].  
  
\bibitem{chandra} 
  S.~Chandrasekhar,
  {\it The Mathematical Theory of Black Holes}
  (Clarendon Press, Oxford, UK, 1998).  
  
\bibitem{mtw} 
  C.~W.~Misner, K.~S.~Thorne and J.~A.~Wheeler,
  {\it Gravitation} (W.~H.~Freeman and Company, San Francisco, California, 1973).  
  
\bibitem{inter}
  M.~Coleman Miller and E.~J.~M.~Colbert,
  Int.\ J.\ Mod.\ Phys.\  D {\bf 13}, 1 (2004)
  [arXiv:astro-ph/0308402].  
  
\bibitem{bh1}
  C.~E.~Rhoades and R.~Ruffini,
  Phys.\ Rev.\ Lett.\  {\bf 32}, 324 (1974).   
  
\bibitem{bh2}
  V.~Kalogera and G.~Baym,
  Astrophys.\ J.\  {\bf 470}, L61 (1996).  
  
\bibitem{bh3} 
  J.~M.~Lattimer,
  Ann.\ Rev.\ Nucl.\ Part.\ Sci.\  {\bf 62}, 485 (2012)
  [arXiv:1305.3510 [nucl-th]].  
 
\bibitem{mass-gap} 
  W.~M.~Farr, N.~Sravan, A.~Cantrell, L.~Kreidberg, C.~D.~Bailyn, I.~Mandel and V.~Kalogera,
  Astrophys.\ J.\  {\bf 741}, 103 (2011)
  [arXiv:1011.1459 [astro-ph.GA]]. 
 
\bibitem{primordial} 
  M.~Y.~Khlopov,
  Res.\ Astron.\ Astrophys.\  {\bf 10}, 495 (2010)
  [arXiv:0801.0116 [astro-ph]]. 
 
\bibitem{bhnum0} 
  E.~P.~J.~van den Heuvel,
  {\it Endpoints of stellar evolution: The incidence of stellar mass black holes in the galaxy},
  in ``Environment Observation and Climate Modelling Through International Space Projects'', 29 (1992). 
 
\bibitem{bhnum} 
  F.~X.~Timmes, S.~E.~Woosley and T.~A.~Weaver,
  Astrophys.\ J.\  {\bf 457}, 834 (1996)
  [astro-ph/9510136].
  
\bibitem{mass} 
  K.~Belczynski, T.~Bulik, C.~L.~Fryer, A.~Ruiter, J.~S.~Vink and J.~R.~Hurley,
  Astrophys.\ J.\  {\bf 714}, 1217 (2010)
  [arXiv:0904.2784 [astro-ph.SR]].  

\bibitem{gg1} 
  A.~Heger and S.~E.~Woosley,
ÊÊAstrophys.\ J.\  {\bf 567}, 532 (2002)
ÊÊ[astro-ph/0107037].

\bibitem{gg2} 
  A.~Heger, C.~L.~Fryer, S.~E.~Woosley, N.~Langer and D.~H.~Hartmann,
ÊÊAstrophys.\ J.\  {\bf 591}, 288 (2003)
ÊÊ[astro-ph/0212469].

\bibitem{gg3} 
  M.~Spera, M.~Mapelli and A.~Bressan,
  Mon.\ Not.\ Roy.\ Astron.\ Soc.\  {\bf 451}, no. 4, 4086 (2015)
  [arXiv:1505.05201 [astro-ph.SR]].

\bibitem{mass2} 
  J.~Casares and P.~G.~Jonker,
  Space Sci.\ Rev.\  {\bf 183}, 223 (2014)
  [arXiv:1311.5118 [astro-ph.HE]].

\bibitem{lm1} 
  L.~R.~Yungelson, J.-P.~Lasota, G.~Nelemans, G.~Dubus, E.~P.~J.~van den Heuvel, J.~Dewi and S.~Portegies Zwart,
  Astron.\ Astrophys.\  {\bf 454}, 559 (2006)
  [astro-ph/0604434].

\bibitem{lm2} 
  P.~D.~Kiel and J.~R.~Hurley,
  Mon.\ Not.\ Roy.\ Astron.\ Soc.\  {\bf 369}, 1152 (2006)
  [astro-ph/0605080].

 \bibitem{lensing} 
  E.~Agol, M.~Kamionkowski, L.~V.~E.~Koopmans and R.~D.~Blandford,
  Astrophys.\ J.\  {\bf 576}, L131 (2002)
  [astro-ph/0203257]. 
  
\bibitem{mass-g} 
  P.~J.~McMillan,
  Mon.\ Not.\ Roy.\ Astron.\ Soc.\  {\bf 465}, 76 (2017)
  [arXiv:1608.00971 [astro-ph.GA]].   
  
\bibitem{mass-bmw} 
  A.~Boehle {\it et al.},
  Astrophys.\ J.\  {\bf 830}, 17 (2016)
  [arXiv:1607.05726 [astro-ph.GA]].   

\bibitem{maoz} 
  E.~Maoz,
  Astrophys.\ J.\  {\bf 494}, L181 (1998)
  [astro-ph/9710309].

\bibitem{postman} 
  M.~Postman {\it et al.},
  Astrophys.\ J.\  {\bf 756}, 159 (2012)
  [arXiv:1205.3839 [astro-ph.CO]].

\bibitem{mv} 
  M.~Volonteri,
  Astron.\ Astrophys.\ Rev.\  {\bf 18}, 279 (2010)
  [arXiv:1003.4404 [astro-ph.CO]]. 

\bibitem{mv07a} 
  M.~Volonteri, G.~Lodato and P.~Natarajan,
  Mon.\ Not.\ Roy.\ Astron.\ Soc.\  {\bf 383}, 1079 (2008)
  [arXiv:0709.0529 [astro-ph]].

\bibitem{mv07b} 
  M.~Volonteri, F.~Haardt and K.~Gultekin,
  Mon.\ Not.\ Roy.\ Astron.\ Soc.\  {\bf 384}, 1387 (2008)
  [arXiv:0710.5770 [astro-ph]].

\bibitem{ferrarese} 
  L.~Ferrarese {\it et al.},
  Astrophys.\ J.\  {\bf 644}, L21 (2006)
  [astro-ph/0603840].

\bibitem{amuse} 
  E.~Gallo, T.~Treu, J.~Jacob, J.~H.~Woo, P.~Marshall and R.~Antonucci,
  Astrophys.\ J.\  {\bf 680}, 154 (2008)
  [arXiv:0711.2073 [astro-ph]].

\bibitem{wu15} 
  X.-B.~Wu {\it et al.},
  Nature\  {\bf 518}, 512 (2015).  

\bibitem{super-e1} 
  P.~Madau, F.~Haardt and M.~Dotti,
  Astrophys.\ J.\  {\bf 784}, L38 (2014)
  [arXiv:1402.6995 [astro-ph.CO]].

\bibitem{super-e2} 
  M.~Bachetti {\it et al.},
  Nature {\bf 514}, 202
  [arXiv:1410.3590 [astro-ph.HE]].

\bibitem{ulxs} 
  E.~J.~M.~Colbert and R.~F.~Mushotzky,
  Astrophys.\ J.\  {\bf 519}, 89 (1999)
  [astro-ph/9901023].

\bibitem{mine} 
  K.~Ohsuga and S.~Mineshige,
  Astrophys.\ J.\  {\bf 736}, 2 (2011)
  [arXiv:1105.5474 [astro-ph.HE]].
  
\bibitem{i-qpo} 
  D.~R.~Pasham, T.~E.~Strohmayer and R.~F.~Mushotzky,
  Nature {\bf 513}, 74 (2014)
  [arXiv:1501.03180 [astro-ph.HE]].  
  
\bibitem{clu1} 
  K.~Gebhardt, R.~M.~Rich and L.~C.~Ho,
  Astrophys.\ J.\  {\bf 578}, L41 (2002)
  [astro-ph/0209313].  
  
\bibitem{clu2} 
  K.~Gebhardt, R.~M.~Rich and L.~C.~Ho,
  Astrophys.\ J.\  {\bf 634}, 1093 (2005)
  [astro-ph/0508251].  
  
\bibitem{fuel1} 
  I.~Shlosman, M.~C.~Begelman and J.~Frank,
ÊÊNature {\bf 345}, 679 (1990).

\bibitem{fuel2} 
  J.~E.~Barnes and L.~Hernquist,
ÊÊAstrophys.\ J.\  {\bf 471}, 115 (1996).

\bibitem{fuel3} 
  L.~Mayer, S.~Kazantzidis, P.~Madau, M.~Colpi, T.~R.~Quinn and J.~Wadsley,
ÊÊScience {\bf 316}, 1874 (2007)
ÊÊ[arXiv:0706.1562 [astro-ph]].  
  
\bibitem{ntm}
  I.~D.~Novikov and K.~S.~Thorne,
  {\it Astrophysics and black holes}, in {\it Black Holes}, edited by C.~De~Witt and B.~De~Witt
  (Gordon and Breach, New York, New York, 1973).  
  
\bibitem{ntm2}
  D.~N.~Page and K.~S.~Thorne,
  Astrophys.\ J.\  {\bf 191}, 499 (1974). 
  
\bibitem{bp1} 
  J.~M.~Bardeen and J.~A.~Petterson,
  Astrophys.\ J.\  {\bf 195}, L65 (1975).
  
\bibitem{bp2}
  S.~Kumar and J.~E.~Pringle,
  Mon.\ Not.\ Roy.\ Astron.\ Soc.\  {\bf 213}, 435 (1985). 
  
\bibitem{bp3} 
  J.~F.~Steiner and J.~E.~McClintock,
  Astrophys.\ J.\  {\bf 745}, 136 (2012)
  [arXiv:1110.6849 [astro-ph.HE]].       
  
\bibitem{thorne} 
  K.~S.~Thorne,
  Astrophys.\ J.\  {\bf 191}, 507 (1974).  
  
\bibitem{sp1} 
  A.~R.~King and U.~Kolb,
  Mon.\ Not.\ Roy.\ Astron.\ Soc.\  {\bf 305}, 654 (1999)
  [astro-ph/9901296].  

\bibitem{sp2} 
  T.~Fragos and J.~E.~McClintock,
  Astrophys.\ J.\  {\bf 800}, 17 (2015)
  [arXiv:1408.2661 [astro-ph.HE]].
  
\bibitem{sp3} 
  E.~Berti and M.~Volonteri,
  Astrophys.\ J.\  {\bf 684}, 822 (2008)
  [arXiv:0802.0025 [astro-ph]].
  
\bibitem{sp4} 
  S.~A.~Hughes and R.~D.~Blandford,
  Astrophys.\ J.\  {\bf 585}, L101 (2003)
  [astro-ph/0208484].  
  
\bibitem{sp5} 
  A.~R.~King and J.~E.~Pringle,
  Mon.\ Not.\ Roy.\ Astron.\ Soc.\  {\bf 373}, L93 (2006)
  [astro-ph/0609598].  
  
\bibitem{sss1} 
  E.~Barausse,
ÊÊMon.\ Not.\ Roy.\ Astron.\ Soc.\  {\bf 423}, 2533 (2012)
ÊÊ[arXiv:1201.5888 [astro-ph.CO]].  
  
\bibitem{sss2} 
  M.~Dotti, M.~Colpi, S.~Pallini, A.~Perego and M.~Volonteri,
ÊÊAstrophys.\ J.\  {\bf 762}, 68 (2013)
ÊÊ[arXiv:1211.4871 [astro-ph.CO]].

\bibitem{sss3} 
  A.~Sesana, E.~Barausse, M.~Dotti and E.~M.~Rossi,
ÊÊAstrophys.\ J.\  {\bf 794}, 104 (2014)
ÊÊ[arXiv:1402.7088 [astro-ph.CO]].

\bibitem{sss4} 
  Y.~Dubois, M.~Volonteri and J.~Silk,
ÊÊMon.\ Not.\ Roy.\ Astron.\ Soc.\  {\bf 440}, no. 2, 1590 (2014)
ÊÊ[arXiv:1304.4583 [astro-ph.CO]].  
  
\bibitem{lamppost} 
  T.~Dauser, J.~Garcia, J.~Wilms, M.~Bock, L.~W.~Brenneman, M.~Falanga, K.~Fukumura and C.~S.~Reynolds,
  Mon.\ Not.\ Roy.\ Astron.\ Soc.\  {\bf 430}, 1694 (2013)
  [arXiv:1301.4922 [astro-ph.HE]].  
  
\bibitem{sandwich} 
  F.~Haardt and L.~Maraschi,
  Astrophys.\ J.\  {\bf 413}, 507 (1993).  
  
 \bibitem{ref} 
  J.~Garcia, T.~Dauser, C.~S.~Reynolds, T.~R.~Kallman, J.~E.~McClintock, J.~Wilms and W.~Eikmann,
  Astrophys.\ J.\  {\bf 768}, 146 (2013)
  [arXiv:1303.2112 [astro-ph.HE]]. 
  
\bibitem{spst} 
  T.~M.~Belloni,
  Lect.\ Notes Phys.\  {\bf 794}, 53 (2010)
  [arXiv:0909.2474 [astro-ph.HE]].  
  
\bibitem{spst2} 
  J.~Homan and T.~Belloni,
  Astrophys.\ Space Sci.\  {\bf 300}, 107 (2005)
  [astro-ph/0412597].  
  
\bibitem{lasota4} 
  J.~P.~Lasota,
  {\it Black hole accretion discs},
  doi:10.1007/978-3-662-52859-4\_1
  arXiv:1505.02172 [astro-ph.HE].  
  
\bibitem{cfm-1915} 
  J.~E.~McClintock, R.~Shafee, R.~Narayan, R.~A.~Remillard, S.~W.~Davis and L.~X.~Li,
  Astrophys.\ J.\  {\bf 652}, 518 (2006)
  [astro-ph/0606076].  

\bibitem{edd1a} 
  R.~F.~Penna, J.~C.~McKinney, R.~Narayan, A.~Tchekhovskoy, R.~Shafee and J.~E.~McClintock,
  Mon.\ Not.\ Roy.\ Astron.\ Soc.\  {\bf 408}, 752 (2010)
  [arXiv:1003.0966 [astro-ph.HE]].  

\bibitem{edd1b} 
  A.~K.~Kulkarni {\it et al.},
  Mon.\ Not.\ Roy.\ Astron.\ Soc.\  {\bf 414}, 1183 (2011)
  [arXiv:1102.0010 [astro-ph.HE]]. 

\bibitem{edd2} 
  J.~F.~Steiner, J.~E.~McClintock, R.~A.~Remillard, L.~Gou, S.~Yamada and R.~Narayan,
  Astrophys.\ J.\  {\bf 718}, L117 (2010)
  [arXiv:1006.5729 [astro-ph.HE]]. 

\bibitem{cfm1} 
  S.~N.~Zhang, W.~Cui and W.~Chen,
  Astrophys.\ J.\  {\bf 482}, L155 (1997)
  [astro-ph/9704072]. 
  
\bibitem{cfm2} 
  L.~X.~Li, E.~R.~Zimmerman, R.~Narayan and J.~E.~McClintock,
  Astrophys.\ J.\ Suppl.\  {\bf 157}, 335 (2005)
  [astro-ph/0411583].   
  
\bibitem{cfm3} 
  J.~E.~McClintock {\it et al.},
  Class.\ Quant.\ Grav.\  {\bf 28}, 114009 (2011)
  [arXiv:1101.0811 [astro-ph.HE]].

\bibitem{cfm4} 
  J.~E.~McClintock, R.~Narayan and J.~F.~Steiner,
  Space Sci.\ Rev.\  {\bf 183}, 295 (2014)
  [arXiv:1303.1583 [astro-ph.HE]].   
  
\bibitem{cfm-1915b} 
  J.~M.~Miller {\it et al.},
  Astrophys.\ J.\  {\bf 775}, L45 (2013)
  [arXiv:1308.4669 [astro-ph.HE]]. 
  
\bibitem{cfm-cyg1} 
  L.~Gou {\it et al.},
  Astrophys.\ J.\  {\bf 742}, 85 (2011)
  [arXiv:1106.3690 [astro-ph.HE]].  
  
\bibitem{cfm-cyg2}
  L.~Gou {\it et al.},
  Astrophys.\ J.\  {\bf 790}, 29 (2014)
  [arXiv:1308.4760 [astro-ph.HE]].    
  
\bibitem{cfm-cyg3}  
  A.~C.~Fabian {\it et al.},
  Mon.\ Not.\ Roy.\ Astron.\ Soc.\  {\bf 424}, 217 (2012)
  [arXiv:1204.5854 [astro-ph.HE]].     
    
\bibitem{cfm-cyg4} 
  J.~A.~Tomsick {\it et al.},
ÊÊAstrophys.\ J.\  {\bf 780}, 78 (2014)
ÊÊ[arXiv:1310.3830 [astro-ph.HE]].  

\bibitem{cfm-cyg5} 
  M.~L.~Parker {\it et al.},
  Astrophys.\ J.\  {\bf 808}, 9 (2015)
  [arXiv:1506.00007 [astro-ph.HE]]. 

\bibitem{cfm-cyg6} 
  D.~J.~Walton {\it et al.},
ÊÊAstrophys.\ J.\  {\bf 826}, 87 (2016)
ÊÊ[arXiv:1605.03966 [astro-ph.HE]].   
  
\bibitem{cfm-gs1354} 
  A.~M.~El-Batal {\it et al.},
ÊÊAstrophys.\ J.\  {\bf 826}, L12 (2016)
ÊÊ[arXiv:1607.00343 [astro-ph.HE]].  
  
\bibitem{cfm-lmcx1} 
  L.~Gou, J.~E.~McClintock, J.~Liu, R.~Narayan, J.~F.~Steiner, R.~A.~Remillard, J.~A.~Orosz and S.~W.~Davis,
  Astrophys.\ J.\  {\bf 701}, 1076 (2009)
  [arXiv:0901.0920 [astro-ph.HE]].   
  
\bibitem{cfm-lmcx1b}   
  J.~F.~Steiner {\it et al.},
  Mon.\ Not.\ Roy.\ Astron.\ Soc.\  {\bf 427}, 2552 (2012)
  [arXiv:1209.3269 [astro-ph.HE]].   
  
\bibitem{cfm-gx339} 
  M.~Kolehmainen and C.~Done,
  Mon.\ Not.\ Roy.\ Astron.\ Soc.\  {\bf 406}, 2206 (2010)
  [arXiv:0911.3281 [astro-ph.HE]].    
  
\bibitem{cfm-gx339b} 
  R.~C.~Reis, A.~C.~Fabian, R.~Ross, G.~Miniutti, J.~M.~Miller and C.~Reynolds,
  Mon.\ Not.\ Roy.\ Astron.\ Soc.\  {\bf 387}, 1489 (2008)
  [arXiv:0804.0238 [astro-ph]].    
  
\bibitem{cfm-gx339c} 
  J.~Garcia {\it et al.},
  Astrophys.\ J.\ {\bf 813}, 84 (2015)
  [arXiv:1505.03607[astro-ph.HE]].   
  
\bibitem{cfm-gx339d} 
  M.~L.~Parker {\it et al.},
ÊÊAstrophys.\ J.\  {\bf 821}, L6 (2016)
ÊÊ[arXiv:1603.03777 [astro-ph.HE]].   
  
\bibitem{cfm-maxi} 
  R.~C.~Reis, J.~M.~Miller, M.~T.~Reynolds, A.~C.~Fabian and D.~J.~Walton,
  Astrophys.\ J.\  {\bf 751}, 34 (2012)
  [arXiv:1111.6665 [astro-ph.HE]].  
  
\bibitem{cfm-liu08} 
  J.~Liu, J.~McClintock, R.~Narayan, S.~Davis and J.~Orosz,
  Astrophys.\ J.\  {\bf 679}, L37 (2008) [Erratum: Astrophys.\ J.\  {\bf 719}, L109 (2010)]
  [arXiv:0803.1834 [astro-ph]].  
  
\bibitem{cfm-sh06} 
  R.~Shafee, J.~E.~McClintock, R.~Narayan, S.~W.~Davis, L.~X.~Li and R.~A.~Remillard,
  Astrophys.\ J.\  {\bf 636}, L113 (2006)
  [astro-ph/0508302].     
  
 \bibitem{cfm-st16}
  J.~F.~Steiner, D.~J.~Walton, J.~A.~Garcia, J.~E.~McClintock, S.~G.~T.~Laycock, M.~J.~Middleton, R.~Barnard and K.~K.~Madsen,
  arXiv:1512.03414 [astro-ph.HE].    

\bibitem{cfm-swift}   
  R.~C.~Reis, A.~C.~Fabian, R.~R.~Ross and J.~M.~Miller,
  Mon.\ Not.\ Roy.\ Astron.\ Soc.\  {\bf 395}, 1257 (2009).
  
\bibitem{cfm-1650} 
  D.~J.~Walton, R.~C.~Reis, E.~M.~Cackett, A.~C.~Fabian and J.~M.~Miller,
  Mon.\ Not.\ Roy.\ Astron.\ Soc.\  {\bf 422}, 2510 (2012)
  [arXiv:1202.5193 [astro-ph.HE]].    
  
\bibitem{cfm-gou_novamus}
  Z.~Chen, L.~Gou, J.~E.~McClintock, J.~F.~Steiner, J.~Wu, W.~Xu, J.~Orosz and Y.~Xiang,
  arXiv:1601.00615 [astro-ph.HE].     
  
\bibitem{cfm-1752} 
  R.~C.~Reis {\it et al.},
  Mon.\ Not.\ Roy.\ Astron.\ Soc.\  {\bf 410}, 2497 (2011)
  [arXiv:1009.1154 [astro-ph.HE]].   
  
\bibitem{cfm-1652} 
  C.~Y.~Chiang, R.~C.~Reis, D.~J.~Walton and A.~C.~Fabian,
  Mon.\ Not.\ Roy.\ Astron.\ Soc.\  {\bf 425}, 2436 (2012)
  [arXiv:1207.0682 [astro-ph.HE]].    
  
\bibitem{cfm-xte} 
  J.~F.~Steiner {\it et al.},
  Mon.\ Not.\ Roy.\ Astron.\ Soc.\  {\bf 416}, 941 (2011)
  [arXiv:1010.1013 [astro-ph.HE]].    
  
\bibitem{cfm-lmcx3} 
  J.~F.~Steiner, J.~E.~McClintock, J.~A.~Orosz, R.~A.~Remillard, C.~D.~Bailyn, M.~Kolehmainen and O.~Straub,
  Astrophys.\ J.\  {\bf 793}, L29 (2014)
  [arXiv:1402.0148 [astro-ph.HE]].    
  
\bibitem{cfm-62} 
  L.~Gou, J.~E.~McClintock, J.~F.~Steiner, R.~Narayan, A.~G.~Cantrell, C.~D.~Bailyn and J.~A.~Orosz,
  Astrophys.\ J.\  {\bf 718}, L122 (2010)
  [arXiv:1002.2211 [astro-ph.HE]].   
  
\bibitem{cfm-h1743} 
  J.~F.~Steiner, J.~E.~McClintock and M.~J.~Reid,
  Astrophys.\ J.\  {\bf 745}, L7 (2012)
  [arXiv:1111.2388 [astro-ph.HE]].   
  
\bibitem{cfm-m31} 
  M.~Middleton, J.~Miller-Jones and R.~Fender,
  Mon.\ Not.\ Roy.\ Astron.\ Soc.\  {\bf 439}, 1740 (2014)
  [arXiv:1401.1829 [astro-ph.HE]].    
  
\bibitem{i1} 
  C.~S.~Reynolds,
  Space Sci.\ Rev.\  {\bf 183}, 277 (2014)
  [arXiv:1302.3260 [astro-ph.HE]]. 

\bibitem{i2} 
  L.~Brenneman,
  {\it Measuring Supermassive Black Hole Spins in Active Galactic Nuclei}
  (Springer New York, 2013)
  [arXiv:1309.6334 [astro-ph.HE]].  
  
\bibitem{i2003} 
  A.~C.~Fabian, K.~Iwasawa, C.~S.~Reynolds and A.~J.~Young,
  Publ.\ Astron.\ Soc.\ Pac.\  {\bf 112}, 1145 (2000)
  [astro-ph/0004366].  
  
\bibitem{suzaku} 
  D.~J.~Walton, E.~Nardini, A.~C.~Fabian, L.~C.~Gallo and R.~C.~Reis,
ÊÊMon.\ Not.\ Roy.\ Astron.\ Soc.\  {\bf 428}, 2901 (2013)
ÊÊ[arXiv:1210.4593 [astro-ph.HE]].  

\bibitem{ngc4051} 
  A.~R.~Patrick, J.~N.~Reeves, D.~Porquet, A.~G.~Markowitz, V.~Braito and A.~P.~Lobban,
  Mon.\ Not.\ Roy.\ Astron.\ Soc.\  {\bf 426}, 2522 (2012)
  [arXiv:1208.1150 [astro-ph.HE]].
  
\bibitem{1h0707} 
  A.~Zoghbi, A.~Fabian, P.~Uttley, G.~Miniutti, L.~Gallo, C.~Reynolds, J.~Miller and G.~Ponti,
  Mon.\ Not.\ Roy.\ Astron.\ Soc.\  {\bf 401}, 2419 (2010)
  [arXiv:0910.0367 [astro-ph.HE]].  
  
\bibitem{ngc3783} 
  L.~W.~Brenneman {\it et al.},
  Astrophys.\ J.\  {\bf 736}, 103 (2011)
  [arXiv:1104.1172 [astro-ph.HE]].  
  
\bibitem{ngc1365a} 
  G.~Risaliti {\it et al.},
ÊÊNature {\bf 494}, 449 (2013)
ÊÊ[arXiv:1302.7002 [astro-ph.HE]].  

\bibitem{ngc1365b} 
  L.~W.~Brenneman, G.~Risaliti, M.~Elvis and E.~Nardini,
  Mon.\ Not.\ Roy.\ Astron.\ Soc.\  {\bf 429}, 2662 (2013)
  [arXiv:1212.0772 [astro-ph.HE]].
  
\bibitem{3c120} 
  A.~M.~Lohfink {\it et al.},
  Astrophys.\ J.\  {\bf 772}, 83 (2013)
  [arXiv:1305.4937 [astro-ph.HE]].  
  
\bibitem{mrk79} 
  L.~C.~Gallo, G.~Miniutti, J.~M.~Miller, L.~W.~Brenneman, A.~C.~Fabian, M.~Guainazzi and C.~S.~Reynolds,
  Mon.\ Not.\ Roy.\ Astron.\ Soc.\  {\bf 411}, 607 (2011)
  [arXiv:1009.2987 [astro-ph.HE]].  
  
\bibitem{mcg63015a} 
  L.~W.~Brenneman and C.~S.~Reynolds,
ÊÊAstrophys.\ J.\  {\bf 652}, 1028 (2006)
ÊÊ[astro-ph/0608502].

\bibitem{mcg63015b} 
  A.~Marinucci {\it et al.},
ÊÊAstrophys.\ J.\  {\bf 787}, 83 (2014)
ÊÊ[arXiv:1404.3561 [astro-ph.HE]].  

\bibitem{iras521} 
  Y.~Tan, J.~Wang, X.~Shu and Y.~Zhou,
  Astrophys.\ J.\  {\bf 747}, L11 (2012)
  [arXiv:1202.0400 [astro-ph.HE]].
  
\bibitem{mrk335} 
  M.~L.~Parker {\it et al.},
ÊÊMon.\ Not.\ Roy.\ Astron.\ Soc.\  {\bf 443}, no. 2, 1723 (2014)
ÊÊ[arXiv:1407.8223 [astro-ph.HE]].  

\bibitem{ark120} 
  E.~Nardini, A.~C.~Fabian, R.~C.~Reis and D.~J.~Walton,
ÊÊMon.\ Not.\ Roy.\ Astron.\ Soc.\  {\bf 410}, 1251 (2011)
ÊÊ[arXiv:1008.2157 [astro-ph.HE]].

\bibitem{swift2127} 
  G.~Miniutti, F.~Panessa, A.~De Rosa, A.~C.~Fabian, A.~Malizia, M.~Molina, J.~M.~Miller and S.~Vaughan,
  Mon.\ Not.\ Roy.\ Astron.\ Soc.\  {\bf 398}, 255 (2009)
  [arXiv:0905.2891 [astro-ph.HE]].

\bibitem{fairall9} 
  S.~Schmoll {\it et al.},
  Astrophys.\ J.\  {\bf 703}, 2171 (2009)
  [arXiv:0908.0013 [astro-ph.HE]].
  
\bibitem{vasudevan} 
  R.~V.~Vasudevan, A.~C.~Fabian, C.~S.~Reynolds, J.~Aird, T.~Dauser and L.~C.~Gallo,
  Mon.\ Not.\ Roy.\ Astron.\ Soc.\  {\bf 458}, 2012 (2016)
  [arXiv:1506.01027 [astro-ph.HE]].  
  
\bibitem{vdk} 
  M.~van der Klis,
  astro-ph/0410551.  
  
\bibitem{motta14} 
  S.~E.~Motta, T.~Munoz-Darias, A.~Sanna, R.~Fender, T.~Belloni and L.~Stella,
  Mon.\ Not.\ Roy.\ Astron.\ Soc.\  {\bf 439}, 65 (2014)
  [arXiv:1312.3114 [astro-ph.HE]].  
  
\bibitem{qqq1} 
  L.~Stella, M.~Vietri and S.~Morsink,
  Astrophys.\ J.\  {\bf 524}, L63 (1999)
  [astro-ph/9907346].

\bibitem{qqq2} 
  M.~A.~Abramowicz and W.~Kluzniak,
  Astron.\ Astrophys.\  {\bf 374}, L19 (2001)
  [astro-ph/0105077].

\bibitem{qqq3} 
  M.~A.~Abramowicz, W.~Kluzniak, Z.~Stuchlik and G.~Torok,
  Astron.\ Astrophys.\  {\bf 436}, 1 (2005)
  [astro-ph/0401464].
  
\bibitem{qpo-smbh} 
  M.~Gierlinski, M.~Middleton, M.~Ward and C.~Done,
  Nature {\bf 455}, 369 (2008)
  [arXiv:0807.1899 [astro-ph]].

\bibitem{qpo-imbh} 
  D.~R.~Pasham, T.~E.~Strohmayer and R.~F.~Mushotzky,
  Nature {\bf 513}, 74 (2014)
  [arXiv:1501.03180 [astro-ph.HE]].  
  
\bibitem{shadow} 
  H.~Falcke, F.~Melia and E.~Agol,
  Astrophys.\ J.\  {\bf 528}, L13 (2000)
  [astro-ph/9912263].  
  
\bibitem{bardeen73}  
  J.~M.~Bardeen, 
  {\it Timelike and null geodesics in the Kerr metric}
  in ``Black Holes'', ed. C.~DeWitt \& B.~S.~DeWitt,
  (Gordon \& Breach, 1973), pp. 215-239.
  
\bibitem{eht}  
  \verb7http://www.eventhorizontelescope.org/7
  
\bibitem{maxim} 
  N.~White,
  Nature {\bf 407}, 146 (2000).  
  
\bibitem{ggww1} 
  E.~E.~Flanagan and S.~A.~Hughes,
ÊÊNew J.\ Phys.\  {\bf 7}, 204 (2005)
ÊÊ[gr-qc/0501041].

\bibitem{ggww2} 
  K.~D.~Kokkotas,
ÊÊRev.\ Mod.\ Astron.\  {\bf 20}, 140 (2008)
ÊÊ[arXiv:0809.1602 [astro-ph]].

\bibitem{ggww3} 
  M.~Maggiore,
  {\it Gravitational Waves: Volume 1: Theory and Experiments}
  (Oxford University Press, 2007).  
  
\bibitem{ligoexp}  
  \verb7http://www.ligo.org7

\bibitem{virgo}
  \verb7http://www.virgo-gw.eu7
  
\bibitem{elisa} 
  P.~A.~Seoane {\it et al.} [eLISA Collaboration],
ÊÊarXiv:1305.5720 [astro-ph.CO].

\bibitem{elisa2} 
  H.~Audley {\it et al.},
ÊÊarXiv:1702.00786 [astro-ph.IM].  

\bibitem{elisa3}  
  \verb7https://www.elisascience.org7
  
\bibitem{decigo}  
  \verb7http://tamago.mtk.nao.ac.jp/decigo/index_E.html7
  
\bibitem{pulsar} 
  G.~Hobbs {\it et al.},
  Class.\ Quant.\ Grav.\  {\bf 27}, 084013 (2010)
  [arXiv:0911.5206 [astro-ph.SR]].  
  
\bibitem{pnm} 
  L.~Blanchet,
  Living Rev.\ Rel.\  {\bf 17}, 2 (2014)
  [arXiv:1310.1528 [gr-qc]].  
  
\bibitem{numrel0} 
  L.~Lehner,
  Class.\ Quant.\ Grav.\  {\bf 18}, R25 (2001)
  [gr-qc/0106072].  
  
\bibitem{numrel1} 
  F.~Pretorius,
ÊÊPhys.\ Rev.\ Lett.\  {\bf 95}, 121101 (2005)
ÊÊ[gr-qc/0507014].

\bibitem{numrel2} 
  M.~Campanelli, C.~O.~Lousto, P.~Marronetti and Y.~Zlochower,
ÊÊPhys.\ Rev.\ Lett.\  {\bf 96}, 111101 (2006)
ÊÊ[gr-qc/0511048].

\bibitem{numrel3} 
  J.~G.~Baker, J.~Centrella, D.~I.~Choi, M.~Koppitz and J.~van Meter,
ÊÊPhys.\ Rev.\ Lett.\  {\bf 96}, 111102 (2006)
ÊÊ[gr-qc/0511103].  
  
\bibitem{teu} 
  S.~A.~Teukolsky,
ÊÊAstrophys.\ J.\  {\bf 185}, 635 (1973).  
  
\bibitem{kkk1} 
  K.~D.~Kokkotas and B.~G.~Schmidt,
  Living Rev.\ Rel.\  {\bf 2}, 2 (1999)
  [gr-qc/9909058].  
  
\bibitem{kkk2} 
  R.~A.~Konoplya and A.~Zhidenko,
  Rev.\ Mod.\ Phys.\  {\bf 83}, 793 (2011)
  [arXiv:1102.4014 [gr-qc]].  
  
\bibitem{bcw} 
  E.~Berti, V.~Cardoso and C.~M.~Will,
  Phys.\ Rev.\ D {\bf 73}, 064030 (2006)
  [gr-qc/0512160].  
  
\bibitem{extp}  
  \verb7http://www.isdc.unige.ch/extp/7
  
\bibitem{snzhang} 
  S.~N.~Zhang {\it et al.} [eXTP Collaboration],
  Proc.\ SPIE Int.\ Soc.\ Opt.\ Eng.\  {\bf 9905}, 99051Q (2016)
  [arXiv:1607.08823 [astro-ph.IM]].   

\bibitem{athena}  
  \verb7http://www.the-athena-x-ray-observatory.eu7
      
\bibitem{review} 
  C.~Bambi,
  Rev.\ Mod.\ Phys.\  {\bf 89}, 025001 (2017)
  [arXiv:1509.03884 [gr-qc]]. 
  
\bibitem{zheng} 
  Z.~Cao, S.~Nampalliwar, C.~Bambi, T.~Dauser and J.~A.~Garcia,
  arXiv:1709.00219 [gr-qc].
  
\bibitem{quantum} 
  S.~B.~Giddings,
  Nature\ Astronomy\ {\bf 1}, 0067 (2017)
  [arXiv:1703.03387 [gr-qc]].     

\end{thebibliography}
\end{document}